\documentclass[aps, prd, onecolumn, tightenlines, notitlepage, superscriptaddress, nofootinbib, preprintnumbers, floatfix,showkeys,11pt]{revtex4-2}

\usepackage[normalem]{ulem}
\usepackage{amstext}
\usepackage{amssymb}
\usepackage{amsmath}
\usepackage{graphicx}
\graphicspath{{figures/}}
\usepackage{url}
\usepackage{ulem}
\usepackage[utf8]{inputenc}
\pdfoutput=1
\usepackage{textcomp}
\usepackage{booktabs,siunitx,array,threeparttable}
\usepackage{gensymb}
\usepackage{enumitem}

\usepackage{epsfig,amsfonts,mathrsfs,amsmath,amssymb,graphicx,slashed,multirow}

\usepackage[x11names]{xcolor}
\usepackage[colorlinks]{hyperref}

\definecolor{seagreen}{rgb}{0.28, 0.25, 0.74}

\usepackage{textgreek} 
\usepackage{caption, subcaption} 
\usepackage{booktabs} 
\usepackage{physics}

\makeatletter
    \newcommand{\colorboxed}[3][white]{\fcolorbox{#2}{#1}{\m@th$\displaystyle#3$}}
\makeatother
\AtBeginDocument{\hypersetup{citecolor=seagreen,linkcolor=seagreen,urlcolor=seagreen}}
\usepackage{appendix}

\begin{document}
\title{{Distinguishing Beyond-Standard Model Effects in Neutrino Oscillation}}

\author{A. Calatayud-Cadenillas}
\email{anthony.calatayud@pucp.edu.pe}
\affiliation{Seccíon Física, Departamento de Ciencias, Pontificia Universidad Cat\'olica del Per\'u, Apartado 1761, Lima, Per\'u}

\author{A. P\'erez-G.}
\email{alicia.perezg@pucp.edu.pe}
\affiliation{Seccíon Física, Departamento de Ciencias, Pontificia Universidad Cat\'olica del Per\'u, Apartado 1761, Lima, Per\'u}

\author{A. M. Gago}
\email{agago@pucp.edu.pe}
\affiliation{Seccíon Física, Departamento de Ciencias, Pontificia Universidad Cat\'olica del Per\'u, Apartado 1761, Lima, Per\'u}

\begin{abstract}
We systematically assess the DUNE experiment’s ability to distinguish between various
beyond-standard neutrino oscillation hypotheses pair combinations. For a pair comparison,
we evaluate the statistical separation, where one hypothesis plays the role of the true signal
while the other corresponds to the test signal. The beyond-standard neutrino oscillation
hypotheses under scrutiny include neutrino decay (invisible and visible), non-standard inter-
actions, quantum decoherence, and the violation of the equivalence principle. We found that the violation of the equivalence
principle is the easiest to differentiate compared to the rest of the hypotheses. Additionally, from our $\chi^2$ statistical separation test, we investigate potential discrepancies between the measured
CP-violation phase $\delta_{CP}$, $\theta_{23}$, and $\Delta m^2_{31}$ relative to their respective true values, which could occur for a given comparison.
In our analysis, we will take the true values of $\delta_{CP}$ as 90$^\circ$ and 180$^\circ$, while $\theta_{23}$ and $\Delta m^2_{31}$ are fixed in its global fit values. In particular, even in
cases where the scenarios of beyond-standard neutrino oscillation hypotheses are statistically indistinguishable, the measured values can exhibit significant deviations from their true values.
\end{abstract}

\maketitle

\section{Introduction}

Neutrino flavor conversion is a well-established phenomenon, supported by extensive experimental evidence spanning nearly three decades~\cite{Cleveland:1998nv, SAGE:1999nng, Super-Kamiokande:2001ljr, SNO:2002tuh, Super-Kamiokande:1998kpq, Kajita:2016vhj, KamLAND:2004mhv, MINOS:2007ixr, MINOS:2013utc}. The data indicate that neutrino flavor oscillation arises from the interference between different neutrino mass eigenstates, a process made possible because the mass eigenstates do not align with the flavor eigenstates~\cite{Kajita:2016vhj,KamLAND:2004mhv, MINOS:2007ixr, MINOS:2013utc}. However, the fact that neutrino oscillation is induced by non-zero and non-equal neutrino masses does not preclude the coexistence of subleading non-standard effects that may be beyond our current experimental sensitivity.

Alternative hypotheses that could play a subleading role in the neutrino oscillation mechanism or modify it include the violation of the equivalence principle by neutrinos~\cite{Gasperini:1988zf,Pantaleone:1992ha}, non-standard interactions of neutrinos with matter~\cite{Valle:1987gv,Guzzo:1991hi}, neutrino decay~\cite{Gelmini:1983ea,Schechter:1981cv}, and quantum decoherence~\cite{Benatti:2000ph,Benatti:2001fa}, among others~\cite{Barger:2000iv,Barbieri:2000mg,Kostelecky:2004xs}. Historically, some of these theoretical hypotheses were competitive solutions to either the solar neutrino problem~\cite{Berezhiani:1992ry,Gago:1999hi} or the atmospheric neutrino anomaly~\cite{Gonzalez-Garcia:1998ryc,Barger:1999bg,Lisi:2000zt}.

Upcoming neutrino long-baseline experiments will not only have enhanced sensitivities for measuring the CP violation phase and solving the neutrino mass hierarchy but also for exploring a new parameter range where these beyond-standard neutrino oscillation(BSO) phenomena may occur~\cite{Hyper-KamiokandeProto-:2015xww,DUNE:2020lwj,DUNE:2020ypp}. Many of these BSO phenomena have been individually studied within these under-development neutrino long-baseline experiments~\cite{Hoefken,Coloma,Masud:2016bvp, Choubey:2017dyu, Coloma:2017zpg, Nosotros, BalieiroGomes:2018gtd, Carpio:2018gum, We-Gabriela-Ternes, Barenboim:2017ewj}. However, if we consider the existence of any BSO phenomena, an interesting question arises that needs further exploration: Shall we be able to distinguish a given BSO mechanism, present in nature, from other exotic phenomena or from the standard neutrino oscillations framework (SO)? Furthermore, how much would the measurement of the CP violation phase, the main objective of these experiments, be distorted if their ability to distinguish among BSO scenarios is not strong enough?.

Our aim in this paper is to answer both questions in the context of the DUNE experiment~\cite{DUNE:2020lwj,DUNE:2020ypp}. This will be achieved by performing a systematic statistical comparison between these BSO scenarios and the SO one. Our goal is to provide a chart containing information about the statistical separation results from these comparisons and to highlight cases where a sizable discrepancy between the real and the measured $\delta_{CP}$ phase occurs. 

For this purpose, we will test the following BSO scenarios: Violation of the Equivalence Principle, Vector Non-Standard Interactions, Neutrino Decay, and Quantum Decoherence. These four BSO scenarios are expected to generate distortions in the extraction of oscillation parameters at DUNE (see, for instance, ~\cite{Hoefken, Coloma, Masud:2016bvp, Choubey:2017dyu, Coloma:2017zpg, Nosotros, BalieiroGomes:2018gtd, Carpio-2, We-Gabriela-Ternes}). This shared characteristic raises the question of whether DUNE will be able to distinguish between these scenarios. Furthermore, some of these scenarios have been proposed as potential solutions to the T2K and NOvA tension ~\cite{NSI-Denton, NSI-Palazzo}. 
A similar study is presented in \cite{Denton:2022pxt}, which focuses in great detail on vector/scalar non-standard neutrino interactions and sterile neutrinos. In this sense, our work is complementary, as it explores theoretical models with different phenomenologies, with vector non-standard interactions being part of the analysis.

This paper is divided as follows: after the introduction, we discuss the theoretical formalism. Next, we present the analysis and simulation details, followed by the results section, and finalize with our conclusions.

\section{Theoretical Formalism}\label{Theoretical-Formalism}

\subsection{Neutrino Oscillation Framework}
The neutrino Hamiltonian in the flavor basis is given by:

\begin{equation}
H^{fv}_{\text{osc}} = \frac{1}{2E_\nu}\left(\mathbf{U} \Delta \mathbf{M}^2 \mathbf{U}^{\dagger}
+ \mathbf{A} \right),
\end{equation}

\noindent where $E_\nu$ represents the neutrino energy, $\mathbf{U}$ is the Pontecorvo-Maki-Nakagawa-Sakata (PMNS) mixing matrix, $\Delta \mathbf{M}^2$ is a diagonal mass matrix defined as $\text{Diag}\left(0, \Delta m^2_{21}, \Delta m^2_{31}\right)$, with $\Delta m^2_{ij} = m^2_{i} - m^2_{j}$. The diagonal matrix $\mathbf{A}$ is given by $\text{Diag}\left( \text{A}_{\text{CC}}, 0, 0 \right)$, encompassing the matter potential $\text{A}_{\text{CC}} = 2\sqrt{2} G_F n_e E_\nu$, where $G_F$ and $n_e$ denote the Fermi constant and the electron number density, respectively. The Hamiltonian above is the Standard Neutrino Oscillation framework in matter; any additional contributions beyond this scheme fall into the Beyond Standard Neutrino Oscillation category. 

\subsection{Beyond Standard Neutrino Oscillation scenarios (BSO)}
As mentioned in the introduction, the four theoretical hypotheses of BSO that we will be working on are the Violation of the Equivalence Principle, Non-Standard Interactions, Neutrino Decay, and Quantum Decoherence. For Non-Standard Interactions (NSI), we will explore two cases: one corresponds to the electron-muon conversion channel (NSI$_{e\mu}$)
and another corresponds to the electron-tau conversion channel (NSI$_{e\tau}$).  Neutrino Decay is subdivided into two sub-cases: Invisible Decay (ID) and Visible Decay (VD) (i.e., Full Decay).

The approach for incorporating the effects of the first two BSO hypotheses, Violation of the Equivalence Principle (VEP) and NSI, into the neutrino oscillation framework is the same. This entails adding a new term to the standard oscillation Hamiltonian, resulting in the following total Hamiltonian ($H^{\text{tot}}_{\text{osc}}$): 

\begin{equation}
H^{\text{tot}}_{\text{osc}}=H^{fv}_{\text{osc}}+H_{\text{BSO}},
\end{equation}

\noindent where $H_{\text{BSO}}$ represents either the VEP contribution or the contribution from NSI. The procedure is quite similar for ID, but with the difference that the term added to $H^{fv}_{\text{osc}}$ is anti-Hermitian. VD considers this, but due to the decay into active neutrinos, other details make calculating the oscillation probability cumbersome, as we will see later. Obtaining the neutrino oscillation probabilities including Quantum Decoherence effects differs from the preceding cases, requiring an open quantum system approach.

\subsubsection{Violation of Equivalence Principle (VEP)}
The Equivalence Principle posits that the gravitational acceleration experienced by an object is independent of its mass, and prompted Einstein to develop the General Theory of Relativity, which establishes the connection between the gravitational field and the curvature of space-time. The Violation of the Equivalence Principle (VEP) can be achieved through the breaking of the universality of Newton’s gravitational constant, $G_N$, being redefined as  
$G_{N}^{'} = \gamma_i G_N$, that implies a mass-dependent gravitational potential $\Phi^{'} = \gamma_i \Phi$, where $\gamma_i$ is a parameter that depends on the mass of the ith particle. 

Under the assumption of the weak field approximation and making the appropriate substitutions and approximations into the space-time metric (see \cite{Hoefken} for details), the following energy-momentum is obtained: 

\begin{equation}
    E_{\nu_i} \simeq p \left(1+2\gamma_i \Phi \right) + \frac{m^2_i}{2 p}.
    \label{Evep}
\end{equation}

\noindent Considering that the momentum $p \simeq E_\nu$,
with $\Delta \gamma_{ij}=\gamma_i -\gamma_j$, this leads to:

\begin{equation}
\Delta E_{ij} = \frac{\Delta m^2_{ij}}{2 E_\nu} + 2 E_\nu \Phi \Delta \gamma_{ij}.
\label{DEvep}
\end{equation}

\noindent The above equation implies that the mass and gravitational eigenstates are coincident bases. However, this might not be the case in a more generic scenario, as the mixing matrix connecting any of these eigenstates to the flavor basis may not be the same. Therefore, the translation of the above equations into a neutrino Hamiltonian in matter in its more generic form goes as follows:  

\begin{equation}
H^{\text{tot}}_{\text{osc}}=H^{fv}_{\text{osc}}+2 E_\nu \Phi \mathbf{U_g}\mathbf{\Delta} \boldsymbol{\gamma}_{\mathbf{ij}} \mathbf{U_g}^{\dagger},
\label{Hvep}
\end{equation}

\noindent $\mathbf{U_g}$ is the mixing matrix that connects 
the gravitational eigenstates with the flavor eigenstates, and 
$\mathbf{\Delta} \boldsymbol{\gamma}_{\textbf{ij}}= \text{Diag}\left(0, \Delta \gamma_{21}, \Delta \gamma_{31}\right)$. For our analysis we 
assume $\mathbf{U_g}=\mathbf{U}$, $\Delta \gamma_{21} > 0 $, and $\Delta \gamma_{31} = 0 $, a particular case of Eq.~(\ref{Hvep}). Our selection of the latter scenario is based on its higher sensitivity to the $\nu_\mu \rightarrow \nu_e$ oscillation channel compared to the alternative scenario, $\Delta \gamma_{21} = 0$ and $\Delta \gamma_{31} \neq 0$ (see Figs. 1 and 2 from~\cite{Hoefken}). We do not consider the case with $\Delta \gamma_{21} \neq 0$, $\Delta \gamma_{31} \neq 0$, and $\mathbf{U_g} \neq \mathbf{U}$ to minimize the number of free parameters in our chosen BSO theories.

\subsubsection{Non-Standard Neutrino Interactions (NSI)}

Since its introduction in \cite{Valle:1987gv,Guzzo:1991hi}, the phenomenon that allows neutrino flavor transitions due to its interaction with matter, called neutrino non-standard interaction with matter (NSI), has been extensively studied \cite{NSI-Gago,NSI-Gago-2,Ohlsson:2012kf}. There are several theories beyond the standard model of physics in which NSI processes are naturally derived (see \cite{Ohlsson:2012kf} and references therein). NSI can occur during neutrino production, propagation, and detection, making it a very rich hypothesis from a phenomenological perspective~\cite{Ohlsson:2012kf,Coloma:2023ixt}. Given the goal of this work, we will only consider NSI during neutrino propagation in matter. With that purpose in mind, it is useful to provide the following effective neutral-current NSI Lagrangian:

\begin{equation}
{\cal{L}}_{eff}^{\text{NSI}}=- 2 \sqrt{2} \epsilon_{\alpha\beta}^{fP} G_F 
\left(\bar{\nu}_{\alpha} \gamma_\mu P_L \nu_\beta \right) \left( \bar{f} \gamma^\mu P f \right), 
\end{equation}

\noindent where $f= e, u, d$, and $\alpha,\beta= e,\mu,\tau$, while $P$ stands for either 
$P_L =\frac{1}{2} \left(1-\gamma_5\right)$ or 
$P_R=\frac{1}{2} \left(1+\gamma_5\right)$.
Now, to write the total Hamiltonian for this case we need to define an effective $\epsilon$ parameter, $\epsilon_{\alpha\beta} = 
\sum_{f,P} \frac{n_f}{n_e} \epsilon^{P}_{\alpha\beta}$, 
with $n_f(n_e)$ are the $f$-type fermion number density. 
Thus, the total Hamiltonian is given by:
\begin{equation}
H^{\text{tot}}_{\text{osc}}=H^{fv}_{\text{osc}}+H_{\text{NSI}},
\end{equation} 

\noindent where:
\begin{equation}
H_{\text{NSI}} = \sqrt{2} G_F n_e 
{
\begin{pmatrix}
\epsilon_{ee} & \epsilon_{e\mu}  & \epsilon_{e\tau} \\
\epsilon^*_{e\mu} & \epsilon_{\mu \mu} & \epsilon_{\mu \tau} \\
\epsilon^*_{e \tau} & \epsilon^*_{\mu \tau} & \epsilon_{\tau \tau}
\end{pmatrix}
}
\end{equation}

Here, the diagonal elements are real and are referred to as the lepton-universality breaking terms. Meanwhile, the non-diagonal elements represent the complex flavor transition terms and are defined as follows:
\begin{equation}
\epsilon_{\alpha\beta} = |\epsilon_{\alpha\beta}| e^{i \phi_{\alpha\beta}}
\end{equation}

In our analysis, we will set all the diagonal elements to zero while turning on only a single non-diagonal parameter, $\epsilon_{\alpha\beta}$, and its complex conjugate, one at a time, with the rest of the parameters set to zero. Consequently, we will work with two different cases: one in which $\epsilon_{e\mu}$ is activated (NSI$_{e\mu}$), and another with $\epsilon_{e\tau}$ (NSI$_{e\tau}$), each being activated individually. The case of 
NSI$_{\mu\tau}$ will not be considered in our analysis since it is indistinguishable from the SO, further details can be seen in section \ref{xi-parameter variation-range-and-BSO-experimental-bounds}. 

\subsubsection{Neutrino decay}
Another way to modify standard neutrino oscillation is by allowing for the possibility that light neutrinos can decay. This hypothesis has been formulated based on the interaction of neutrinos with a massless scalar particle called the Majoron under two types of couplings, the scalar $g_s$ and the pseudoscalar $g_p$ (see \cite{Nosotros} and references therein). These interactions are described by the Lagrangian below:

\begin{equation}
    {\cal{L}}_{int}=\frac{(g_s)_{ij}}{2} \bar{\nu}_i \nu_j J 
    + i \frac{(g_p)_{ij}}{2} \bar{\nu}_i \gamma_5 \nu_j J 
\end{equation}

Any of these terms can lead to neutrino decay such as $\nu_j\rightarrow \nu_i +J$, where $J$ is the Majoron, and the $\nu_j$, $\nu_i$ are the $j,i$ neutrino-mass states. In our analysis, only the scalar coupling will be considered. Depending either on the ability to observe the final state neutrinos or whether they are active or not, neutrino decay can be classified in two ways:

\begin{enumerate}[label=(\alph*)]
\item{\bf Invisible decay (ID)}
In this scenario, the parent neutrino can decay either into a sterile neutrino or into an active neutrino that is not observable, due to its energy being below the experimental threshold. This leads to a modification in the total neutrino matter Hamiltonian, as described by the equation:

\begin{equation}
H^{\text{tot}}_{\text{osc}}=H^{fv}_{\text{osc}} - \frac{i}{2}\mathbf{U} \Gamma \mathbf{U}^{\dagger}
\label{Htotdecay}
\end{equation}

For the present analysis, we will use $\mathbf{\Gamma}= \text{Diag}\left(0, 0, \Gamma_{3}\right)$ for normal hierarchy, where $\Gamma_{3}$ represents the decay rate of $\nu_3 \rightarrow \nu_x$. For inverted hierarchy, we consider $\mathbf{\Gamma}= \text{Diag}\left(0, \Gamma_{2}, 0\right)$ for inverted hierarchy, where $\Gamma_{2}$ represents the decay rate of $\nu_2 \rightarrow \nu_x$.
The term $- \frac{i}{2}\mathbf{U} \Gamma \mathbf{U}^{\dagger}$ is anti-Hermitian, 
with $\mathbf{U}$ being the PMNS matrix. Given that the $H^{\text{tot}}_{\text{osc}}$, due to the introduction of the anti-hermitian term, became an anti-hermitian one, this matrix  will be diagonalized by a non-unitary matrix $\widetilde{\mathbf{U}}$ in the following way: 

\begin{equation}
\label{eq:decaymatterdiagon}
\widetilde{\mathbf{U}}^{-1} H^{\text{tot}}_{\text{osc}} \widetilde{\mathbf{U}} =\text{Diag} \left(\widetilde{\lambda}_1, \tilde{\lambda}_2, \widetilde{\lambda}_3 \right)
\end{equation}
with 
\begin{equation}
 \widetilde{\lambda}_J = \frac{\widetilde{m}_J^2}{2 E_\nu} -i\frac{\widetilde{\Gamma}_J}{2}
\end{equation}

It is important to note that despite considering a single decay rate in the mass eigenstates basis, when we go to the effective matter eigenstates with the inclusion of matter effects, we can find non-zero values for all decay rates in this diagonalized basis. Therefore, the neutrino oscillation probability is given by~\cite{Nosotros}: 

\begin{equation}
\label{probNuDcy}
P_{\text{inv}}(\nu_\alpha \rightarrow \nu_\beta)=
\left|\sum_{J} \widetilde{U}_{\alpha J}^{-1}
\exp\left[-i\frac{\widetilde{m}^2_J L}{2 E_\alpha}\right] \exp\left[-\frac{\widetilde{\alpha}_J L}{2 E_\alpha}\right]
\widetilde{U}_{J \beta} \right|^2
\end{equation}

where $J=1,2,3$ and $\widetilde{\Gamma} _J
= \dfrac{\widetilde{\alpha}_J}{E_\alpha}$, is 
the alternative parameter we will use to characterize neutrino decay from now on.

\item{\bf Visible decay (Full decay)}
The neutrino visible decay (VD) occurs when the decay products are (observable) active neutrinos with lower energies than their parent neutrinos. The full decay (FD) transition probability \cite{Nosotros}, which contains neutrino ID and VD contributions, goes as follows:   

\begin{equation}
    P_{\text{dec}}\left(\nu_\alpha^{(r)} \rightarrow \nu_\beta^{(s)} \right) =
     P_{\text{inv}}\left(\nu_\alpha^{(r)} \rightarrow \nu_\beta^{(s)} \right) \delta_{rs} \delta\left(E_\alpha -E_\beta \right) + P_{\text{vis}}\left(E_\alpha, E_\beta \right)
\end{equation}

\noindent where the $r=(+,-)$ and $s=(+,-)$ are the helicities. The first term is the ID contribution (with 
the correspondence $\widetilde{U}_{\alpha j} \rightarrow 
\widetilde{U}_{\alpha j}^{(r)}$ 
and $\widetilde{U}_{j \beta} \rightarrow 
\widetilde{U}^{(s)}_{j \beta})
$, which cannot change neutrino-helicities $(r,s)$ during either the conversion or the survival. In addition, we must note that the neutrino mixing (in general) is related to the helicities as follows: 
$\widetilde{U}^{(-)} = \widetilde{U}$ and 
$\widetilde{U}^{(+)} = \widetilde{U}^{*}$. Besides, in the ID the $\nu_\alpha$ and $\nu_\beta$ must have the same energy. To ensure the aforementioned conditions, the $\delta_{rs}$ and $\delta\left(E_\alpha -E_\beta \right)$ are added. The second term is the VD contribution, which, as expected, implies a degradation of the decaying neutrino energy $E_\alpha$, higher than the $E_\beta$. With minor modifications, here we are following the prescription for the $P_{\text{vis}}$ in \cite{Nosotros}, which considers that the neutrino decays only once in its path:

\begin{equation}
\begin{split}
\label{eq:pvis_basic}
P_{\rm vis}(E_\alpha,\,E_\beta)&= \int d\ell\,\left|\sum_{J}\left(\widetilde {U}^{(r)}\right)^{-1}_{\alpha J} 
\exp\left[-i\frac{\widetilde {m}_J^2\, \ell}{2E_\alpha}\right]\exp\left[-
\frac{\widetilde{\alpha}_J\, \ell}{2 E_\alpha}\right]
\sum_{i=2}^3\sum_{j=1}^{i-1}\widetilde{C}^{(r)}_{Ji}\right. \\
&\left.\times\sqrt{\frac{d}{dE_\beta}\Gamma_{\nu_i^r\to\nu_j^s}(E_\alpha)}\sum_{K}\left(\widetilde{C}^{(s)}\right)_{jK}^{-1}\exp\left[-i\frac{\widetilde{m}_K^2 (L-\ell)}{2E_\beta}\right]\right.\\
&\left.\times
\exp\left[-\frac{\widetilde {\alpha}_K (L-\ell)}{2 E_\beta}\right]\widetilde{U}_{K\beta}^{(s)}\right|^2
\end{split}
\end{equation}

\noindent with $J,K=1,2,3$. It is important to note that $\widetilde {U}^{(r)}_{\alpha J}$, $\widetilde {m}_J^2$, $\widetilde{\alpha}_J$ and 
$\widetilde{C}^{(r)}_{Ji}$ are calculated using Eq.~(\ref{eq:decaymatterdiagon}) with $E_\nu = E_\alpha$, while 
 $\widetilde {U}^{(s)}_{K \beta}$, $\widetilde {m}_K^2$, $\widetilde{\alpha}_K$ and $\widetilde{C}^{(s)}_{Kj}$ are obtained for 
 $E_\nu = E_\beta$. The reasoning behind the structure of $P_{\rm vis}$ is as follows: we initiate with a  flavor (interaction)
 eigenstate $\nu_\alpha$, which is transformed into the matter eigenstate $\widetilde{\nu}_J$ before propagating a distance $l$ prior to decay. In the decay process $\nu_i^r\to\nu_j^s$, we initially rotate from the $\widetilde{\nu}_J$ to the mass eigenstate $\nu_i^r$ (where the decay is defined) using the operator $\widetilde{C}^{(r)}_{Ji}$. After the decay, we revert the decay product, the mass eigenstate $\nu_j^s$, back to the matter eigenstate $\widetilde{\nu}_K$ with the inverse of the operator $\widetilde{C}^{(s)}_{Kj}$. The $\widetilde{\nu}_K$ then propagates until it reaches the detector, where it is rotated into the flavor eigenstate $\nu_\beta$. The definition of the operator 
 $\widetilde{C}$ is:
 
 \begin{equation}
\widetilde{C}^{(h)}_{Mn}
=\sum_{\alpha =e, \mu, \tau} \widetilde {U}^{(h)}_{\alpha M}
\left({U} \right)^{(h)*}_{\alpha n}
 \end{equation}
 
\noindent where $U$ is a PMNS matrix element, $h=(+,-)$ is the helicity, and $M,n$ is a matter and mass eigenstate, respectively.

For FD, we will use $\alpha^{vis}_3 = \alpha_{\nu_3 \rightarrow \nu_1}$ for normal ordering (NO), referred to as FD1, and $\alpha^{vis}_2 = \alpha_{\nu_2 \rightarrow \nu_3}$ for inverted ordering (IO), referred to as FD2. We take $m_1 =0.05$ eV and a scalar coupling for both cases. It is worth mentioning that the strength of the neutrino decay, $\alpha$, is largely unaffected by the choice of $m_1$, as shown in Figs. 4 and 6 from~\cite{Bound-FD}. Here, we adopt the scalar coupling, as it offers slightly better sensitivity than the pseudoscalar coupling for the analysis presented in the following sections.   

\end{enumerate}

\subsubsection{Quantum decoherence (QD)}

The neutrino system can be envisioned as an open quantum system subjected to the effects of its interaction with the environment~\cite{Benatti:2000ph,Benatti:2001fa}, which can be a manifestation of Planck-scale physics~\cite{Barenboim:2006xt,Mavromatos:2004sz}. These effects are typically revealed by introducing quantum decoherence (QD) parameters in the neutrino oscillation probabilities. However, it is also possible to have CPT violation effects~\cite{Carrasco:2018sca,Capolupo:2018hrp} or even be sensitive to the Majorana CP violation phase~\cite{Buoninfante:2020iyr,Carrasco-Martinez:2020mlg}, among others.  

\begin{enumerate}[label=(\alph*)]

\item{\bf The density matrix formalism}

The way to describe the evolution of the neutrino system in a dissipative environment is by using the Lindblad master equation, which implies representing the neutrino system with the density matrix formalism. The Lindblad master equation is defined as follows~\cite{Carpio-2}: 

\begin{equation}
\frac{d\rho(t)}{dt} = -i \left[H,\rho(t)\right] + L\left[\rho(t)\right]
\label{lindblad}
\end{equation}

where $H$ is the neutrino system Hamiltonian (that we will discuss later) and $\rho(t)$ is the neutrino density matrix, while $L\left[\rho(t)\right]$ is
the term that encloses the 
dissipative effects. The $L\left[\rho(t)\right]$ is given by~\cite{We-Gabriela-Ternes}:

\begin{equation}
L\left[\rho(t)\right] = -\frac{1}{2} \sum_{j} \left\lbrace A^{\dagger}_{j} A_{j} \rho(t) +
\rho(t) A^{\dagger}_{j} A_{j} \right\rbrace
+\sum_{j} A_{j} \rho(t) A^{\dagger}_{j}
\end{equation}

The $A_{j}$ are a set of operators with $j=1,2,...,8$, for three-neutrino generations. These operators are hermitian to secure the increase of the Von Neumann entropy over time. Another step for solving the Eq.~(\ref{lindblad}) is to write the operators $\rho$, $H$ and $A_j$ as: 

\begin{equation}
\rho =\frac{1}{2}\sum{\rho_\mu \lambda_\mu}, \,\, \,\, H =\frac{1}{2}\sum{h_\mu \lambda_\mu},\,\,\,\, A_j=\frac{1}{2}\sum{a^j_\mu \lambda_\mu}
\end{equation} 

\noindent where $\mu$ is running from 0 to 9, $\lambda_0$ is the identity matrix and $\lambda_k$ are the Gell- Mann matrices, which satisfy the relationship $\left[ \lambda_i,\lambda_j \right]=2 i f_{ijk} \lambda_k$, the $f_{ijk}$ is the structure constant of $SU(3)$. The matrix $L_{\mu\nu} (\equiv L[\rho(t)])$  is real and symmetric, and it can be expressed as $L_{kj}=\frac{1}{2} \sum_{l,m,n}{(\vec{a}_{n} . \vec{a}_{l}) f_{knm} f_{mlj}} $, all of this due to the Hermiticity of the $A_j$, with the components $L_{\mu0}=L_{0\mu}=0$, because of the probability conservation. Additionally, the decoherence matrix $\textbf{L} \equiv L_{kj}$ must satisfy the Cauchy-Schwartz inequalities and the complete positivity condition, to ensure that the eigenvalues of $\rho(t)$ are positive at any time. Taking all the above into consideration, Eq.~(\ref{lindblad}) turns out to be: 

\begin{equation}
\dot{\rho}_0=0, \,\,\,\, \dot{\rho}_k=\left( H_{kl}+ L_{kl} \right) \rho_{l} = M_{kl} \rho_{l}
\label{masterComp}
\end{equation}
        \noindent with $H_{kl}=\sum_{j} h_j f_{jlk}$. Therefore, the solution of $\rho(t)$ 
is: 

\begin{equation}
\varrho(t) = e^{\textbf{M}t} \varrho(0)
\end{equation}
where $\varrho$ is an eight columns vector and $\textbf{M}\equiv M_{kl}$. The $e^{\textbf{M}t}$ is written as:
\begin{equation}
e^{\textbf{M}t} = \textbf{D} e^{\textbf{M}_{D} t} \textbf{D}^{-1}
\equiv \left[ e^{\textbf{M}t} \right]_{il} = \sum_k D_{il} e^{\lambda_k} D^{-1}_{kl}
\end{equation}

\noindent with $\textbf{M}_{D} =\textbf{Diag} \left(\lambda_1,.....,\lambda_8\right)$. Thus, the neutrino oscillation probability is 
given by:

\begin{equation}
P_{\nu_\alpha \rightarrow \nu_\beta} = \text{Tr}\left( \rho^\alpha(t)\rho^\beta \right) 
= \frac{1}{3}+\frac{1}{2} \sum_{i,j} \rho^{\beta}_i(0) 
\left[ e^{\textbf{M}t} \right]_{il} \rho^{\alpha}_l(0)
=\frac{1}{3}+\frac{1}{2} (\varrho^\beta(0))^T\varrho^\alpha(t)
\end{equation}

In this work the decoherence matrix is given by: 
$\textbf{L}=-\text{Diag}(0,0,0,\Gamma,\Gamma,\Gamma,\Gamma,0)$ where the experimental bounds for $\Gamma$ are well determined~\cite{Bound-QD}. It is worth noting that QD disrupts the coherence pattern, resulting in damping factors as $e^{-\Gamma L}$ multiplying the oscillatory terms contained in the neutrino flavor transition
probability.

\item{\bf Neutrino Hamiltonian and 
rotation between quantum bases}

We assume that decoherence effects are defined in the neutrino mass eigenstate basis, where the neutrino Hamiltonian is described as:
\begin{equation}
H^{ms}_{\text{osc}} = \mathbf{U}^{\dagger} H^{fv}_{\text{osc}} \mathbf{U}
\end{equation}

To solve Eq. (\ref{masterComp}), we are following the recipe given in \cite{Carpio-2}. Thus, we must transform the aforementioned equation from the mass eigenstate basis to the matter eigenstate basis, where the Hamiltonian is diagonal and more manageable. For this purpose, we apply the unitary transformation: $\mathbf{U}^{\dagger}_T O_v \mathbf{U}_T$, where $O_v$ can represent $\rho$, $H$, or $L\left[\rho(t)\right]$, defined in the mass eigenstate basis. The matrix $\mathbf{U}_T$ facilitates the rotation between matter eigenstates and mass eigenstates and is defined as $\mathbf{U}_T = \mathbf{U}^{\dagger} \mathbf{U}_m$, where the matrix $\mathbf{U}_m$ is responsible for diagonalizing $H^{fv}_{\text{osc}}$.
\end{enumerate}

\section{Analysis and Simulation details}
\subsection{BSO quantification parameter \texorpdfstring{$\xi$}{\texttwoinferior}}\label{BSO-quantification-parameter}

To perform direct comparisons between the different BSO scenarios, we need to define a singular/equivalent parameter that quantifies the magnitude of the BSO effect relative to the standard one. We will denote this BSO strength parameter as $\xi$. For the cases of VEP and NSI, the identification of $\xi$ would be done by considering the oscillatory term $\frac{\Delta m^2}{4E}L \sim \mathcal{O}(1)$. Therefore, as can be derived from Eq. (\ref{DEvep}), the BSO oscillatory term from VEP that perturbs its standard counterpart is $E_\nu \Phi \Delta \gamma_{ij}L$. Said perturbation is the $\xi$ parameter for VEP. Similarly, for NSI, the corresponding $\xi$ parameter is equal to $2\sqrt{2} G_F n_e |\epsilon_{\alpha\beta}| L$. 

In the case of invisible neutrino decay, the size of its effect in distorting the SO formula is expressed through the magnitude of the
 exponential decay factor $e^{-\frac{\alpha_3}{E_\nu}L}$ for NO or $e^{-\frac{\alpha_2}{E_\nu}L}$ for IO. Thus, the corresponding BSO strength parameter for invisible decay is $\xi =\left(\alpha_3/E_\nu\right)L$ for NO or $\xi =\left(\alpha_2/E_\nu\right)L$ for IO. Meanwhile, in the case of visible decay or FD, the BSO strength parameter can be either
 $\xi =\left(\alpha^{vis}_3/E_\nu \right) L$ for NO (FD1) 
 or $\xi =\left(\alpha^{vis}_2/E_\nu \right) L$ for IO (FD2). As we mentioned earlier, damping factor $e^{-\Gamma L}$ characterizes the effects of QD in the neutrino oscillation probability, which is a factor that attenuates the oscillatory behavior of the latter. Therefore, as expected, for quantum decoherence, the $\xi =\Gamma L$. 
All the definitions of $\xi$ for the different scenarios we study here are summarized in Table \ref{epsdef}, where for DUNE
$L=1284.9$ km, $\langle E_\nu \rangle = 2.6$ GeV and $\rho = 2.848$~g/cm$^3$~\cite{DUNE:2020ypp}. The relationship between $n_e$ and $\rho$ is $\rho = n_e/(Y_e N_A)$, where $N_A$ is the Avogadro constant and assuming the electron fraction $Y_e=0.5$. 

 \begin{table}[htbp]
    \centering
    \begin{tabular}{|c|c|}
        \hline
        \textbf{BSO scenario} & \textbf{Strength parameter $\xi$} \\ \hline
        \textbf{VEP} & $\langle E_\nu \rangle \Phi \Delta \gamma_{21}L$ \\ 
        \textbf{NSI}$_{e\mu}$ &
        $2\sqrt{2} G_F n_e |\epsilon_{e\mu}|L$ \\ 
        \textbf{NSI}$_{e\tau}$ &
        $2\sqrt{2} G_F n_e |\epsilon_{e\tau}|L$ \\
        \textbf{ID (NO)} & $\dfrac{\alpha_3 L}{\langle E_\nu \rangle} $ \\ 
        \textbf{ID (IO)}& $\dfrac{\alpha_2 L}{\langle E_\nu \rangle} $ \\ 
        \textbf{FD (NO)} & $\dfrac{\alpha^{vis}_3 L}{\langle E_\nu \rangle}$ \\
        \textbf{FD (IO)} & $\dfrac{\alpha^{vis}_2 L}{\langle E_\nu \rangle}$\\
        \textbf{QD} & $\Gamma L$ \\ \hline
    \end{tabular}
    \caption{Definition of BSO strength parameter, $\xi$, for all scenarios considered. $L$ and $\langle E_\nu \rangle$ are the experiment baseline and average energy, respectively, in our case are the ones defined for the DUNE experiment.}
     \label{epsdef}
\end{table}

\begin{table}[htbp]
    \centering
    \begin{tabular}{|c|c|c|c|}
        \hline
        \textbf{BSO scenario} & \textbf{BSO  parameter} &  \textbf{Experimental upper bound} & \textbf{value at  $\xi_{\text{max}}=0.05$}\\ 
       \hline
        \textbf{VEP} & $\Phi \Delta \gamma_{21}$  & $6.97 \times 10^{-23}$~\cite{Bound-VEP} & $2.95 \times 10^{-24}$ \\
        \textbf{NSI}$_{e\mu}$ & $|\epsilon_{e\mu}|$ & $4.55 \times 10^{-2}$~\cite{Coloma:2023ixt} & $3.53 \times 10^{-2}$ \\
        \textbf{NSI}$_{e\tau}$ & $|\epsilon_{e\tau}|$ & $1.67 \times 10^{-1}$~\cite{Coloma:2023ixt} & $3.53 \times 10^{-2}$ \\
        \textbf{ID (NO)} & $\alpha_3 $ & $2.40 \times 10^{-4}$ eV$^2$~\cite{Bound-ID} & $2.00 \times 10^{-5}$ eV$^2$ \\ 
        \textbf{ID (IO)} & $\alpha_2 $ & $2.40 \times 10^{-4}$ eV$^2$~\cite{Bound-ID} & $2.00 \times 10^{-5}$ eV$^2$ \\ 
       \textbf{FD (NO)} & $\alpha^{vis}_3  $  & $7.80 \times 10^{-5}$ eV$^2$~\cite{Bound-FD} & $2.00 \times 10^{-5}$ eV$^2$ \\ 
        \textbf{FD (IO)} & $\alpha^{vis}_2  $  & $7.80 \times 10^{-5}$ eV$^2$~ \cite{Bound-FD} & 
        $2.00 \times 10^{-5}$ eV$^2$\\ 
        \textbf{QD} & $\Gamma $ & $4.80 \times 10^{-23}$ GeV~\cite{Bound-QD} & $7.68 \times 10^{-24}$ GeV  \\ 
        \hline
    \end{tabular}  
    \caption{Experimental upper bounds for each BSO characteristic parameter and its value attained for the maximum $\xi_{\text{max}}=0.05$. The experimental bounds for VEP and NSI, are build from~\cite{Bound-VEP} and~\cite{Coloma:2023ixt}, the details are given in the text.}
     \label{ExpBounds}
\end{table}
\subsection{ \texorpdfstring{$\xi$} - parameter variation range and BSO experimental bounds} \label{xi-parameter variation-range-and-BSO-experimental-bounds}
We will generate different sets of simulated true data for various BSO scenarios, each corresponding to a specific $\xi$ parameter value within its defined range, with a maximum of 0.05. The correspondence between $\xi = 0.05$ and each BSO scenario characteristic parameter used for generating its true simulated data is displayed in Table~\ref{ExpBounds}. 

In Table~\ref{ExpBounds} are also displayed the experimental bounds for the different BSO scenarios. These bounds will act as upper limits, constraining the fitting range of the BSO scenarios when they play the role of the theoretical hypothesis. Constraints on the NSI parameters $\epsilon_{e\mu}$ and $\epsilon_{e\tau}$ were obtained from \cite{Coloma:2023ixt}. 

To maintain a conservative yet not overly restrictive approach, we consider the average of the strongest bounds given in Table 5 in \cite{Coloma:2023ixt}. We average the strongest limit from the 90\% confidence level intervals and 99\% confidence level, both taken in absolute value as calculated in the analysis that incorporates NSI in electron scattering and CE$\nu$NS (see \cite{Coloma:2023ixt} and references therein).

On the other hand, as mentioned earlier, 
we will not consider NSI$_{\mu\tau}$ as a BSO true scenario since the IceCube limit on $|\epsilon_{\mu\tau}| \sim {\cal{O}}(10^{-3})$~\cite{IceCube:2022ubv} is beyond the sensitivity reach of DUNE. As a result, this scenario will not yield meaningful results in terms of distinguishing it from other BSO hypotheses (which serve as theoretical benchmarks) or showing noticeable distortions in the measurement of 
$\delta_{CP}$, $\theta_{23}$ and $\Delta m^2_{31}$ relative to their true values.


The experimental constraint for the VEP parameter, i.e., for $\Phi \Delta \gamma_{21}$, is translated from the Lorentz Invariance Violation bound
for $|a_{e \mu}|$ given
at 90\% C.L., and obtained in~\cite{Bound-VEP} using NO$\nu$A~\cite{NOvA:2021nfi,NOvA:2023iam}. For this translation, we use Eq. (18) from \cite{Hoefken}, which defines the relation between $\Phi \Delta \gamma_{21}$ and $|a_{e \mu}|$ (equivalent to $\tilde{\nu}_{e \mu}$ in~\cite{Hoefken}). It is worth mentioning that the resulting experimental constraint is one order of magnitude above the value corresponding to the maximum \(\xi\) used to simulate events with VEP (see Table \ref{ExpBounds}). In that sense, we can see that, in general, the BSO experimental bounds in Table~\ref{ExpBounds} are above the 
maximum value that can be achieved for the true BSO parameters.

In Table~\ref{TrductorXitoBSM}, we display the translation between the characteristic parameters of the different BSO scenarios and different values of the intensity parameter $\xi$, within our analysis range. 
\begin{table}[htbp]
    \centering
    \begin{tabular}{|c|c|c|c|c|c|c|c|c|}
        \hline
        \textbf{$\xi$} 
        & $\Phi \Delta \gamma_{21}$[$10^{-24}$]
        & $|\epsilon_{e\mu}|$, $|\epsilon_{e\tau}|[10^{-2}]$
        & $\alpha_3$, $\alpha_2$, $\alpha^{vis}_3$, $\alpha^{vis}_2$ [$10^{-5}$ eV$^2$]
        &$\Gamma $ [$10^{-24}$ GeV]
        \\ 
       \hline
        0.005 & $0.30 $ & $0.35$ & $0.20$ & $0.77$ 
        \\
        0.010 & $0.60$ & $0.70$ & $0.40$ & $1.54 $ 
        \\
        0.020 & $1.18$ & $1.41$ & $0.80$ & $3.07$ 
        \\
        0.030 & $1.77$ & $2.12$ & $1.20$ & $4.61$ 
        \\
        0.040 & $2.36$ & $2.83$ & $1.60$ & $6.14$ 
        \\
        0.050 & $2.95$ & $3.53$ & $2.00$ & $7.68$ 
        \\
        \hline
    \end{tabular}  
    \caption{Equivalency between the BSO characteristic parameters and different values of $\xi$, that will be part of our analysis range.}
     \label{TrductorXitoBSM}
\end{table}

\subsection{Simulation details}

The DUNE experiment~\cite{DUNE:2020lwj,DUNE:2020ypp}, whose main goals are to achieve precise measurements of $\delta_{CP}$ and the neutrino mass ordering, is a long-baseline neutrino experiment currently under construction. DUNE consists of two detectors: the near detector and the far detector. The near detector will be located at ${\cal{O}}(100 \text{m})$ from the neutrino beam source, while the far detector, to be placed at 1284.9 km, will have 40 kt of fiducial-volume liquid Argon. These detectors will be impacted by a neutrino beam sent from Fermilab toward the Sanford Underground Research Laboratory (the far detector location), after traveling through the Earth with an average matter density of 2.848 $\text{g/cm}^3$. The neutrino beam will be produced from the collisions of 120 GeV protons with a beam power of 1.2 MW (i.e., 624 kt-MW-years of exposure), operating for 6.5 years in neutrino mode (FHC) and 6.5 years in antineutrino mode (RHC), the time periods that we will use in our simulation. Furthermore, for our analysis, we will take into account $\nu_e, \bar{\nu}_e$ appearance and $\nu_\mu, \bar{\nu}_\mu$ disappearance channels, following, for all purposes, the prescription given~\cite{We-Gabriela-Ternes}.

While the implementation of DUNE in GLoBES has been given by the collaboration and is widely used, BSO phenomena, such as VEP and NSI, have to be introduced into the package through already existing functions (registering new probability engines) or by changing the source code directly (modifying the hamiltonian defined in the package to be diagonalized). 

More complex theoretical hypotheses such as ID, FD (in which the Hamiltonian is non-Hermitian), and QD (with the density matrix formalism) were computed by programming the probability functions externally and integrating them into the GLoBES event rates calculations. For QD, the SQuIDS \cite{SQuIDS} library was specifically used to calculate the probability quickly and efficiently.

\begin{table}[]
    \centering
    \begin{tabular}{|c|c|c|}
        \hline
         \textbf{Standard parameter} & \textbf{Value for NO} & \textbf{Value for IO} \\ \hline
         $\theta_{12}$[$^\circ$] & 34.3 & 34.3 \\ 
         $\theta_{13}$[$^\circ$] & 8.53 & 8.58 \\ 
         $\theta_{23}$[$^\circ$] & 49.26 & 49.46 \\
         $\Delta m^2_{21}$[$10^{-5}$eV$^{2}$] & 7.50 & 7.50 \\
         $\Delta m^2_{31}$[$10^{-3}$eV$^{2}$] & 2.55 & -2.45 \\
         \hline
    \end{tabular}
    \caption{Neutrino standard oscillation parameters values (taken from \cite{deSalas:2020pgw}) fixed for the generation of all true BSO scenarios. }
    \label{SOparams}
\end{table}

The neutrino oscillation parameters given in in Table \ref{SOparams} will be fixed along the generation of the BSO true simulated events, for all the different scenarios. For the simulation of the NSI true events, regardless of the case, we decided to fix the corresponding complex phase of the NSI parameters at $-90^{\circ}$, inspired by~\cite{NSI-Palazzo,NSI-Denton}.

\subsection{Statistical analysis} \label{Statistical-analysis}
Our analysis strategy consists of generating a set of true simulated BSO data for different values of $\xi$, within a range between $[0,0.05]$, and for a given true BSO scenario. Then, for each value of $\xi$, we perform a statistical comparison between the given BSO true scenario and each of the remaining BSO theoretical hypotheses, plus SO (the test scenario). Our $\chi^2$ statistical test is defined for 
a fixed $\xi$ as follows:
\begin{equation}
\begin{split}
\label{eq:chi2}
 \chi^2_\xi(\vec{\kappa}) =& \min_{\vec{\alpha}} \sum_{\text{ch}} \left\lbrace 2\sum_i \left[ N^{\text{test}}_{\text{ch},i}(\vec{\kappa},\vec{\alpha}) - N^{\text{true}}_{\text{ch},i}(\xi) +
 N^{\text{true}}_{\text{ch},i} (\xi) \ln \left(\frac{N^{\text{true}}_{\text{ch},i}(\xi)}{N^{\text{test}}_{\text{ch},i}(\vec{\kappa},\vec{\alpha})}\right)\right] 
\right\rbrace + \sum_j \left(\frac{\alpha_{j}}{\sigma_{\alpha_{j}}}\right)^2 \\
& + \left(\frac{\theta^{\text{true}}_{13}-\theta^{\text{test}}_{13}}{\sigma_{\theta_{13}} }\right)^2 ,
\end{split}
\end{equation}
the first term is the summation of the appearance channels, 
$\nu_e$ and $\bar{\nu}_e$, and the disappearance ones, $\nu_\mu$ and $\bar{\nu}_\mu$, identified by the index 
$\text{ch}$, $i$ represents the 
number of each energy bin. As expected, the $N^{\text{true}}_{\text{ch},i}(\xi)$ corresponds to the BSO true event rates for 
fixed $\xi$ and a given hypothesis, while the 
$N^{\text{test}}_{\text{ch},i}(\vec{\kappa},\vec{\alpha})$  are the test events that could correspond to any of the other BSO or the SO hypotheses, depending on which scenario is under test. The $N^{\text{test}}_{\text{ch},i}(\vec{\kappa},\vec{\alpha})$ depends on 
$\vec{\kappa} =(\vec{\zeta},\delta_{CP},\theta_{23},\Delta m^2_{31})$ where, in most BSO scenarios, $\vec{\zeta}$ is equal to the single characteristic parameter of the given BSO test hypothesis, with the exception of NSI,  where $\vec{\zeta}=(|\epsilon_{\alpha \beta}|,\phi_{\alpha \beta})$. It is evident that for SO, working as a test hypothesis, $\vec{\zeta}=0$. 

The BSO parameter and $\delta_{CP}$ are free, while we include a $3\sigma$ prior for $\theta_{13}$ for NO(IO), with $\sigma_{\theta_{13}} \simeq 0.38^\circ (0.39^\circ)$ \cite{deSalas:2020pgw}, this is the third term of the equation above. Therefore, for our statistical analysis, we consider that VEP, ID, FD, and QD test hypotheses have 
two degrees of freedom (d.o.f), NSI and SO have three and one d.o.f, respectively. To take into account the systematic uncertainties, the 
$\vec{\alpha}$ nuisance parameters are included as normalization factors of the signal and background components of a given channel $\text{ch}$; the standard deviations $\vec{\sigma}_{\alpha_j}$ are provided by the DUNE files for GLoBES~\cite{DUNE:2021cuw}. The second term includes the penalty factors due to the systematic uncertainties
$\alpha_j$ with the standard deviation 
$\sigma_{\alpha_j}$ and zero as expectation value. 

Our $\chi^2$ represents the minimization in $\vec{\alpha}$ nuisance parameters for
a fixed $\vec{\kappa}$. Therefore, the best-fit point is obtained by minimizing over \(\kappa\):
\begin{equation}
    \chi^{2}_{\xi, \text{min}} = \min_{\vec{\kappa}} \left\lbrace \chi^2_\xi (\vec{\kappa})\right\rbrace = \chi^2_\xi ( \vec{\zeta}^{fit}, \delta_{CP}^{fit}, \theta_{23}^{fit},
    \Delta m^2_{31}{}^{fit}) ,
    \label{eq:chi2min}
\end{equation}
where $\vec{\zeta}^{fit}$, $\delta_{CP}^{fit}$, $\theta_{23}^{fit}$, and 
$\Delta m^2_{31}{}^{fit}$
are the best fit parameters. Since our goal is to estimate the 
the statistical separation between BSO scenarios (true compared to test), and between BSO true scenario and SO as test, we will have to use the $\Delta \chi^2$:
\begin{equation}
\Delta \chi^2 =\chi^{2}_{\xi, \text{min}} -\chi^{2}_{\xi, \text{min(true=test)}}
\label{deltachi2}
\end{equation}
where the $\chi^{2}_{\xi, \text{min(true=test)}}$ is equal to zero and correspond to the case when 
the BSO true and the test scenario are the same.  


\subsection{Plots description and guidelines}

Our findings will be presented through four different types of plots:
\begin{itemize}
    \item {\bf $N_\sigma$ versus $\xi$}: This plot measures, for a given value of $\xi$, the $N_\sigma$ separation between our true hypothesis and the fitted one, taking into consideration the number of d.o.f, where 1 d.o.f is used for SO, 3 d.o.f for the 
NSI scenarios and 2 d.o.f for the rest of BSO scenarios.

   \item {\bf $\delta_{CP}$ versus $\xi$, $\theta_{23}$ versus $\xi$, and 
   $\Delta^2_{31}$ versus $\xi$}:  These plots shows the 
   behavior of the $\delta_{CP}$, 
   $\theta_{23}$ and $\Delta^2_{31}$  fitted values as function of $\xi$. Different levels of uncertainties of the latter are displayed to assess the degree of discrepancy between their
fitted values and the true ones. The uncertainties on the
   $\delta_{CP}$, 
   $\theta_{23}$ and $\Delta^2_{31}$ true values are 
   determined for a given $\xi$ value (i.e. according to their corresponding 
\end{itemize}

At follows we present our guidelines for the plot presentation:
\begin{enumerate}
 \item \textbf{Largest $N_\sigma$ value}: We will display only plots $N_\sigma$ versus $\xi$ plots in which one or more of BSO or SO theoretical hypotheses tested reach 
 an $N_\sigma > 3 \sigma$ separation relative to the corresponding BSO true hypothesis. 
  \item \textbf{The fitted $\delta_{CP}$, $\theta_{23}$, and 
$\Delta m^2_{31}$ reach at least the 2.4$\sigma$ level of deviation from their corresponding true parameter uncertainties}: 
We will present plots of $\delta_{CP}$ vs. $\xi$, $\theta_{23}$ vs. $\xi$, and 
$\Delta m^2_{31}$ vs. $\xi$, where one or more of the fitted $\delta_{CP}$ curves, derived from testing a specific BSO theoretical hypothesis as true, exceed the $2.4\sigma$ uncertainty level of their corresponding true parameter. As previously noted, these uncertainties are determined for each value of $\xi$.

    \item \textbf{All fitted BSO parameters are within the experimental bounds}: 
    All the BSO fitted parameters that construct the BSO curves displayed in the plots are below the experimental bounds. When a fitted BSO parameter exceeds the experimental bounds, it is not included in the curve. These discarded fitted points correspond to the discontinuities or blank regions in the BSO curves, which partially cover the $\xi$ range.
    
    \item \textbf{BSO trivial solutions}: 
    To ensure that the effects of our fit scenarios are detectable by DUNE's sensitivity, we have established a lower fitting bound for each BSO parameter. This is a minimum value that they must reach to be considered a non-trivial solution, meaning it does not overlap with the SO scenario and significantly differs from it. This bound would correspond to take $\xi = 0.005$,
    one order of magnitude less than our strongest BSO true effect ($\xi = 0.05$). The specific BSO parameters that result in a trivial solution are displayed in Table~\ref{LowBounds}.

\begin{table}[htbp]
    \centering
    \begin{tabular}{|c|c|c|c|c|c|c|c|c|}
        \hline
        \textbf{BSO parameter} 
        & $\Phi \Delta \gamma_{21}$ 
        & $|\epsilon_{e\mu}|$, $|\epsilon_{e\tau}|$
        & $\alpha_3$, $\alpha_2$, $\alpha^{vis}_3$, $\alpha^{vis}_2$ [eV$^2$]
        &$\Gamma $ [GeV]
        \\ 
       \hline
        \textbf{Lower fitting bound} & $2.95 \times 10^{-25}$ & $3.53 \times 10^{-3}$ & $2.00 \times 10^{-6}$ & $7.68 \times 10^{-25}$ 
        \\
        \hline
    \end{tabular}  
    \caption{Lower fitting bounds for each BSO characteristic parameter. All the fitted parameters 
    below these bounds are considered a 
    BSO trivial solution.}
     \label{LowBounds}
\end{table}

\end{enumerate}

\section{Results}
We will present our results separated by each BSO true scenario. As mentioned earlier, the $|\epsilon_{\mu\tau}|$ scenario will not be included because there is no plot of either $N_\sigma$ versus $\xi$  or $\delta_{CP}$ versus $\xi$ for this scenario that fulfills our criteria 1 (for $N_\sigma$) or 2 (the fitted $\delta_{CP}$).

\subsection{\textbf{VEP}}

\begin{figure}
        \vspace{-0.8cm}
	\centering
	\begin{subfigure}[t]{0.99\textwidth}
		\centering
		\includegraphics[width=\linewidth]{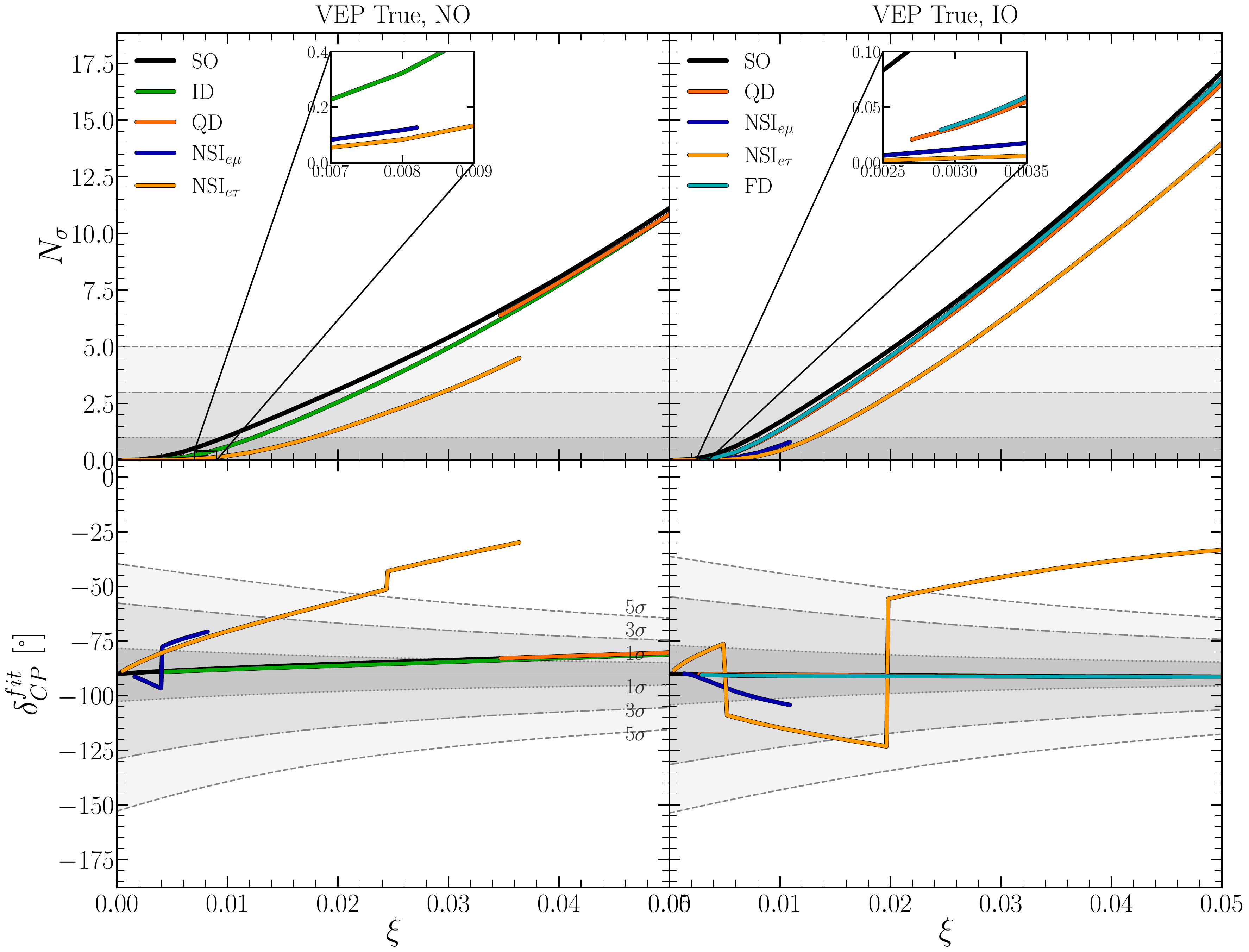}
	\end{subfigure}

        \vspace{-0.3cm}
	\caption{VEP as a True Model. At the top is N$_{\sigma}$ vs $\xi$ and at the bottom is $\delta_{CP}^{fit}$ vs $\xi$. All plots were made with $\delta_{CP}^{true} = -90^\circ$.}
 \label{A-Plots_VEP_dcp_-90}
\end{figure}

In Fig.~\ref{A-Plots_VEP_dcp_-90}, we display the plots of $N_\sigma$ vs $\xi$ and $\delta_{CP}$ vs $\xi$ for both NO and IO, with $\delta^{true}_{CP}=-90^\circ$, since criteria 1 and 2 from our guidelines are satisfied. For NO, we see that FD is not plotted as a theoretical hypothetical solution because, within the whole range
of $\xi$, it falls into the BSO trivial solution category. This means that the corresponding fitted parameters for FD are effectively zero (falling below the limits specified in Table.~\ref{LowBounds}), coinciding with the SO solution. For IO, we can observe that ID is the BSO trivial. On the other hand, a common characteristic of theoretical curves is that some of them may partially cover the entire range of $\xi$, showing discontinuities either at the beginning or at the end of the curve. For example, if they start at $\xi>0$, it implies that the missing part of the curve corresponds to the BSO trivial solution category. Conversely, if the curves end at 
$\xi<0.05$ means that beyond this endpoint, the fitted parameters are excluded by experimental bounds. 
In the case where VEP, assuming NO, is taken as the true solution, the QD and ID (NSI$_{e\mu}$, and NSI$_{e\tau}$) theoretical hypotheses exhibit beginning(ending) discontinuities. For IO, QD, and FD (NSI$_{e\mu}$) correspond to the beginning (ending) discontinuity cases.
The $N_\sigma=5\sigma$ separation between the different theoretical hypotheses and the simulated true data of VEP begins at $\xi = 0.028$ and $0.030$ for SO and ID, while QD begins at $N_\sigma>5\sigma$ , when NO is assumed. For IO the $N_\sigma=5\sigma$ begins at $\xi=0.020, 0.021, 0.022$ and $0.027$, for SO, FD, QD and NSI$_{e\tau}$. 

Regarding the discrepancy between $\delta^{true}_{CP}=-90^\circ$ and the fitted $\delta^{fit}_{CP}$, we observe that for NO, a distortion of
3(5)$\sigma$ is only achieved for NSI$_{e\tau}$ when $\delta^{fit}_{CP} = -64.54^\circ(-53.85^\circ)$ for $\xi=0.014 (0.022)$. It is expected to observe big distortions for NSI$_{e\tau}$ since, in this scenario, the complex phase $\phi_{e\tau}$ represents an additional d.o.f of variation. For this case SO, ID, and QD never reach the $3 \sigma$. Similarly to the NO case, for IO, NSI$_{e\tau}$ shows a 3(5)$\sigma$ discrepancy at $\xi=0.015 (0.023)$ with $\delta^{fit}_{CP} = -119.81^\circ(-52.52^\circ)$. For NSI$_{e\tau}$, there is a degenerate behavior which has already been pointed out in~\cite{Liao:2016hsa}. In fact, for NO, we have a less dramatic degeneracy for the case of NSI$_{e\mu}$ and other almost negligible for NSI$_{e\tau}$. As expected, these abrupt changes in $\delta^{fit}_{CP}$ are correlated with those appearing in the other mixing parameters, which are being fitted simultaneously. For the other scenarios, SO, QD, and FD, there are no distortions. 

\begin{figure}
        \vspace{-0.8cm}
	\centering
	\begin{subfigure}[t]{0.99\textwidth}
		\centering
		\includegraphics[width=\linewidth]{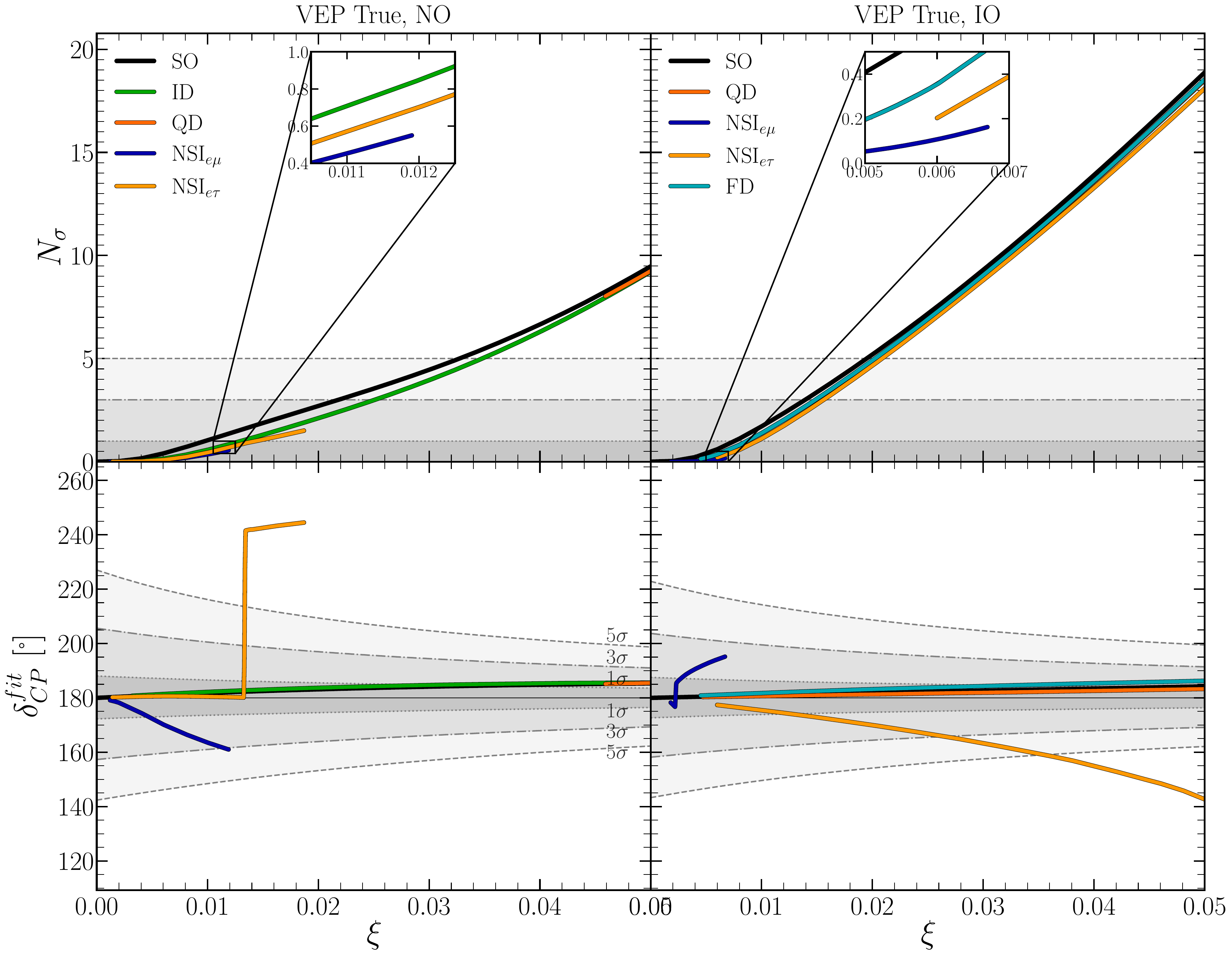}
	\end{subfigure}

        \vspace{-0.3cm}
	\caption{VEP as a True Model. At the top is N$_{\sigma}$ vs $\xi$ and at the bottom is $\delta_{CP}^{fit}$ vs $\xi$. All plots were made with $\delta_{CP}^{true} = 180^\circ$.}
 \label{A-Plots_VEP_dcp_180}
\end{figure}

In Fig.~\ref{A-Plots_VEP_dcp_180} we show the plots corresponding to $\delta^{true}_{CP}=180^\circ$. As we can observe, for the latter case and NO, FD is the only 
BSO trivial solution. Meanwhile, for IO, ID is the only BSO trivial solution. For NO, a steep change occurs at $\xi=0.013$, abruptly achieving a much higher separation than $5\sigma$, with $\delta^{fit}_{CP}=241.49^\circ$, which is significantly far from $180^\circ$. In the case of IO, the $N_\sigma=5\sigma$ separation is reached at $\xi=0.035$, with $\delta^{fit}_{CP}=159.07^\circ$, for $\text{NSI}_{e\tau}$. Furthermore, for NO, $3\sigma$ is reached for NSI$_{e\mu}$ at $\xi = 0.011$ with $\delta^{fit}_{CP} = 161.54^\circ$.

\begin{figure}
        \vspace{-0.8cm}
	\centering
	\begin{subfigure}[t]{0.99\textwidth}
		\centering
		\includegraphics[width=\linewidth]{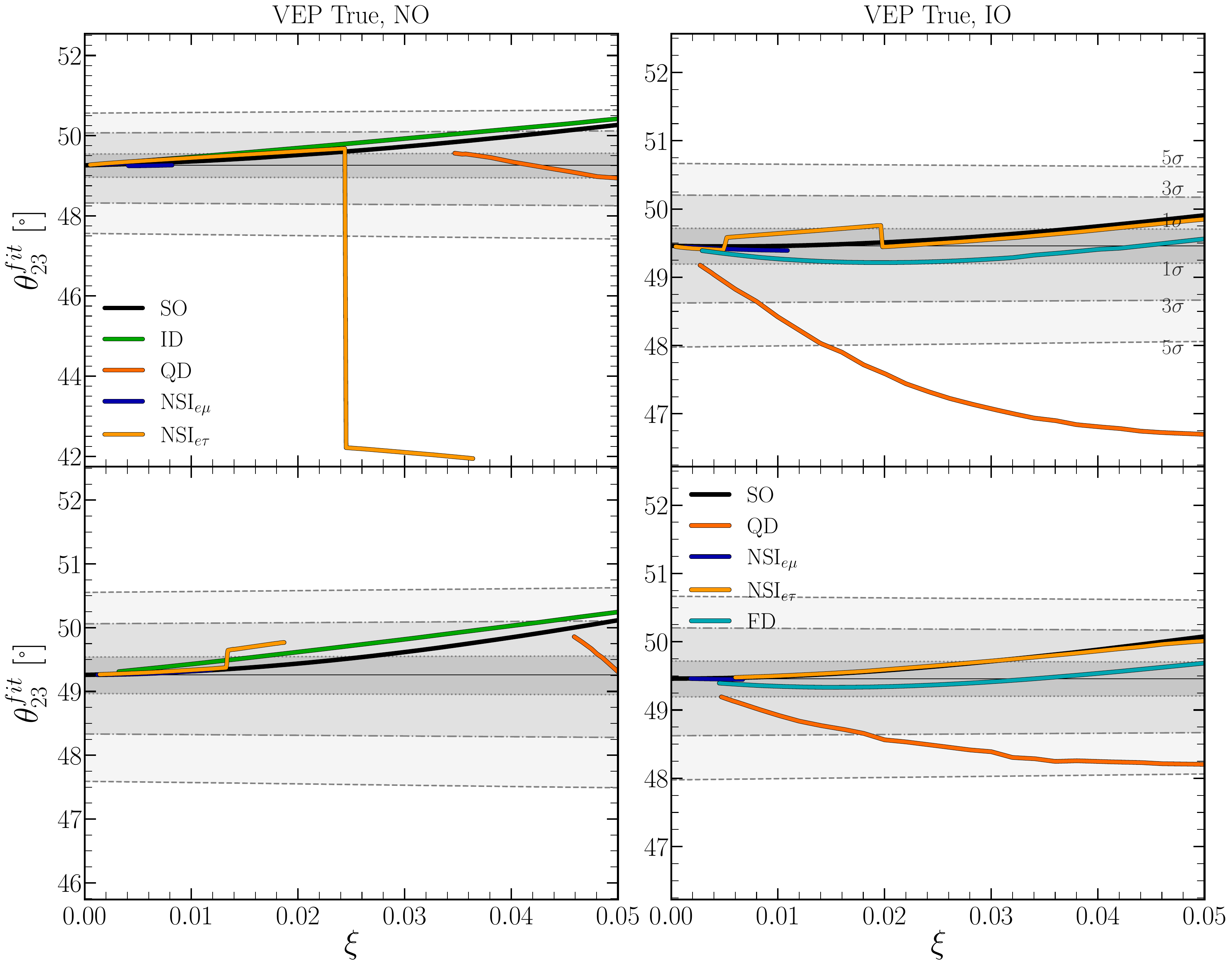}
	\end{subfigure}

        \vspace{-0.3cm}
	\caption{VEP as a True Model. Plots at the top was made with $\delta_{CP}^{true} = -90^\circ$ and iat the bottom with $\delta_{CP}^{true} = 180^\circ$. }
 \label{A-Plots_VEP_th23_-90_180}
\end{figure}

In Fig.~\ref{A-Plots_VEP_th23_-90_180} we display the plots $\theta^{fit}_{23}$ versus $\xi$. Here, for NO and $\delta^{true}_{CP} = -90^\circ$, we observe that for NSI$_{e\tau}$, $\theta^{fit}_{23}$ significantly exceeds the 5$\sigma$ discrepancy at $\xi = 0.025$ through a steep change in its trend, reaching $\theta^{fit}_{23} = 42.21^\circ$. As anticipated, this change occurs at the same value of $\xi = 0.025$ as for $\delta^{fit}_{CP}$. This is not surprising, as it is expected that all the different fitted mixing parameters are correlated at the same value of $\xi$. On the other hand, SO and ID reach the 3$\sigma$ discrepancy at $\xi = 0.045$ and $\xi = 0.037$, respectively. The remaining scenarios do not attain the 3$\sigma$ discrepancy. For IO and $\delta^{true}_{CP} = -90^\circ$, the QD scenario exceeds the 5$\sigma$ distortion at $\xi = 0.014$, while the others remain below the 3$\sigma$ threshold. For NO and $\delta^{true}_{CP} = 180^\circ$, only SO and ID achieve the 3$\sigma$ distortion at $\xi \sim 0.05$ and $\xi = 0.043$, respectively. For IO and $\delta^{true}_{CP} = 180^\circ$, only the QD scenario surpasses the 3$\sigma$ discrepancy at $\xi = 0.018$.
 
\begin{figure}
        \vspace{-0.8cm}
	\centering
	\begin{subfigure}[t]{0.99\textwidth}
		\centering
		\includegraphics[width=\linewidth]{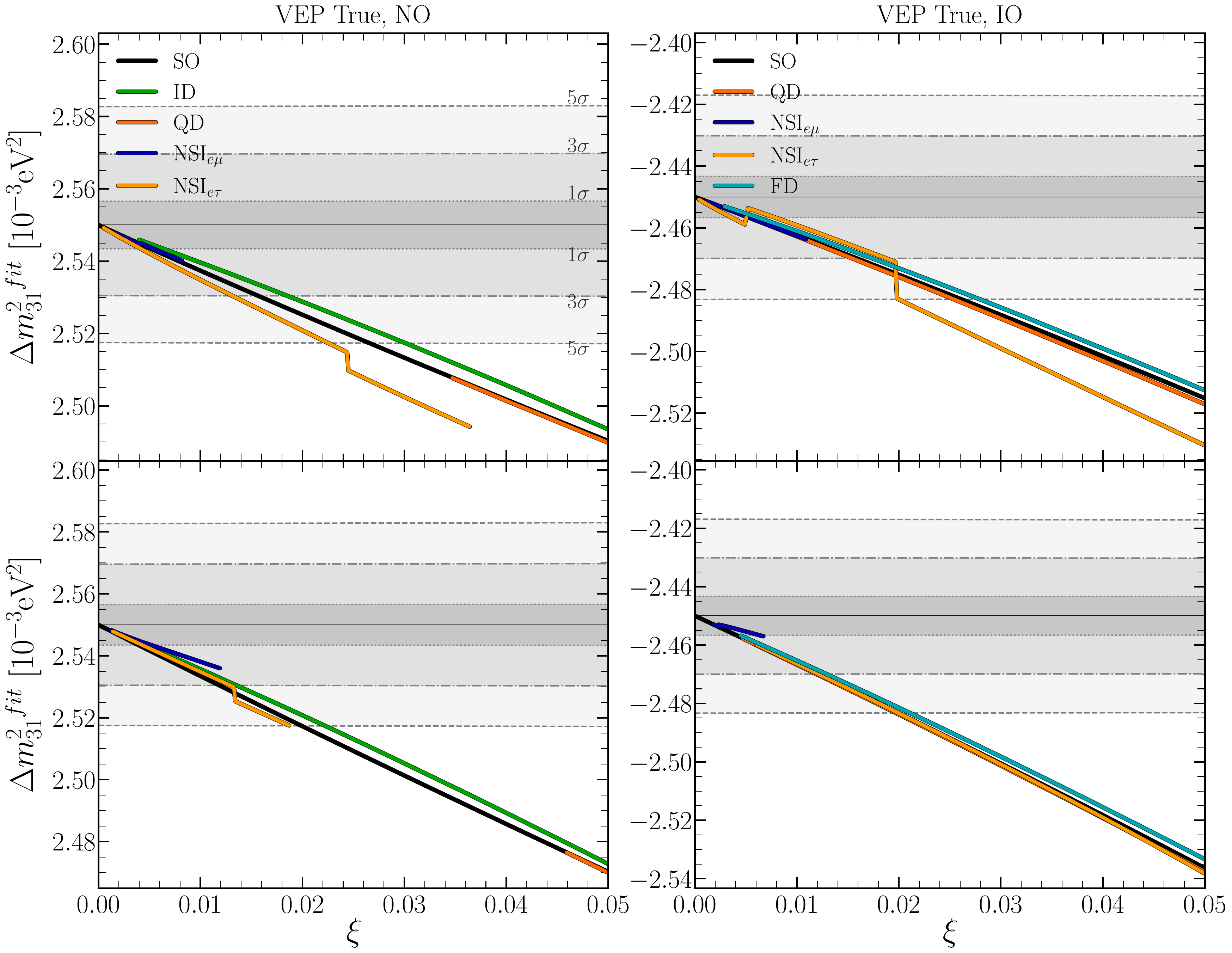}
	\end{subfigure}

        \vspace{-0.3cm}
	\caption{VEP as a True Model. Plots at the top was made with $\delta_{CP}^{true} = -90^\circ$ and iat the bottom with $\delta_{CP}^{true} = 180^\circ$.}
 \label{A-Plots_VEP_dcp_-90_180_dm31}
\end{figure}

In Fig.~\ref{A-Plots_VEP_dcp_-90_180_dm31}, we display the plots of
$\Delta m^2_{31}{}^{fit}$. For NO and 
$\delta^{true}_{CP} = -90^\circ$, SO, ID and 
NSI$_{e\tau}$ reach the $5\sigma$ separation at $\xi=0.027, 0.030$ and $0.023$, respectively and with $\Delta m^2_{31}{}^{fit}=2.517\times10^{-3}\text{eV}^2$. The QD scenario begins for a $\sigma$ deviation higher than 5$\sigma$. As we predicted, the steep change for $\text{NSI}_{e\tau}$ occurs at $\xi=0.025$, the same value of $\xi$ as for the other fitted mixing parameters. For IO, SO, QD, $\text{NSI}_{e\tau}$ and FD achieve the 5$\sigma$ separation for $\xi=0.026,0.025,0.020$ and $0.028$, respectively, with
$\Delta m^2_{31} {}^{fit}=-2.483\times10^{-3}\text{eV}^2$. For NO and 
$\delta^{true}_{CP} = 180^\circ$, SO and ID reach 
the 5$\sigma$ deviation at 
$\xi=0.020$ and $0.022$, with
$\Delta m^2_{31}{}^{fit}=2.517\times10^{-3}\text{eV}^2$. The QD scenario, as in the case of $\delta^{true}_{CP} =-90^\circ$, begins at
$\xi=0.046$ corresponding to a much higher value than the
5$\sigma$ deviation. The  $\text{NSI}_{e\tau}$ exceeds the 3$\sigma$ separation ending almost at 5$\sigma$. For IO, SO, QD, $\text{NSI}_{e\tau}$ and FD, surpass the 5$\sigma$  at $\xi=0.020$ and $0.021$(FD) with $\Delta m^2_{31}{}^{fit}=-2.483\times10^{-3}\text{eV}^2$.

\subsection{\textbf{NSI}}
\begin{figure}
        \vspace{-0.8cm}
	\centering
	\begin{subfigure}[t]{1.00\textwidth}
		\centering
		\includegraphics[width=\linewidth]{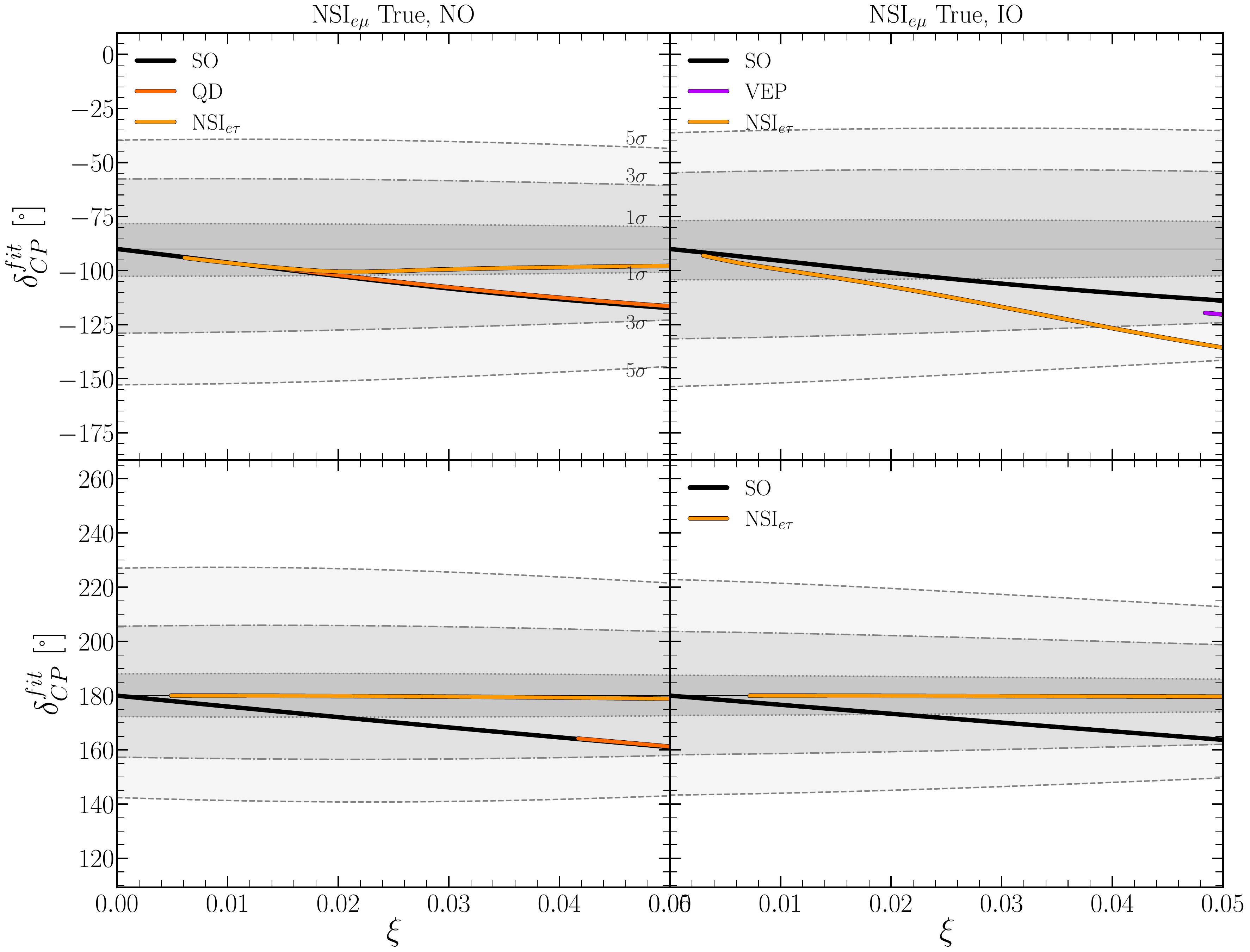}
	\end{subfigure}

        \vspace{-0.3cm}
	\caption{NSI$_{e\mu}$ as a True Model. Plots at the top was made with $\delta_{CP}^{true} = -90^\circ$ and at the bottom with $\delta_{CP}^{true} = 180^\circ$.}
 \label{A-Plots_NSI_EM_dcp_-90}
\end{figure}

\begin{figure}
        \vspace{-0.8cm}
	\centering
	\begin{subfigure}[t]{0.50\textwidth}
		\centering
		\includegraphics[width=\linewidth]{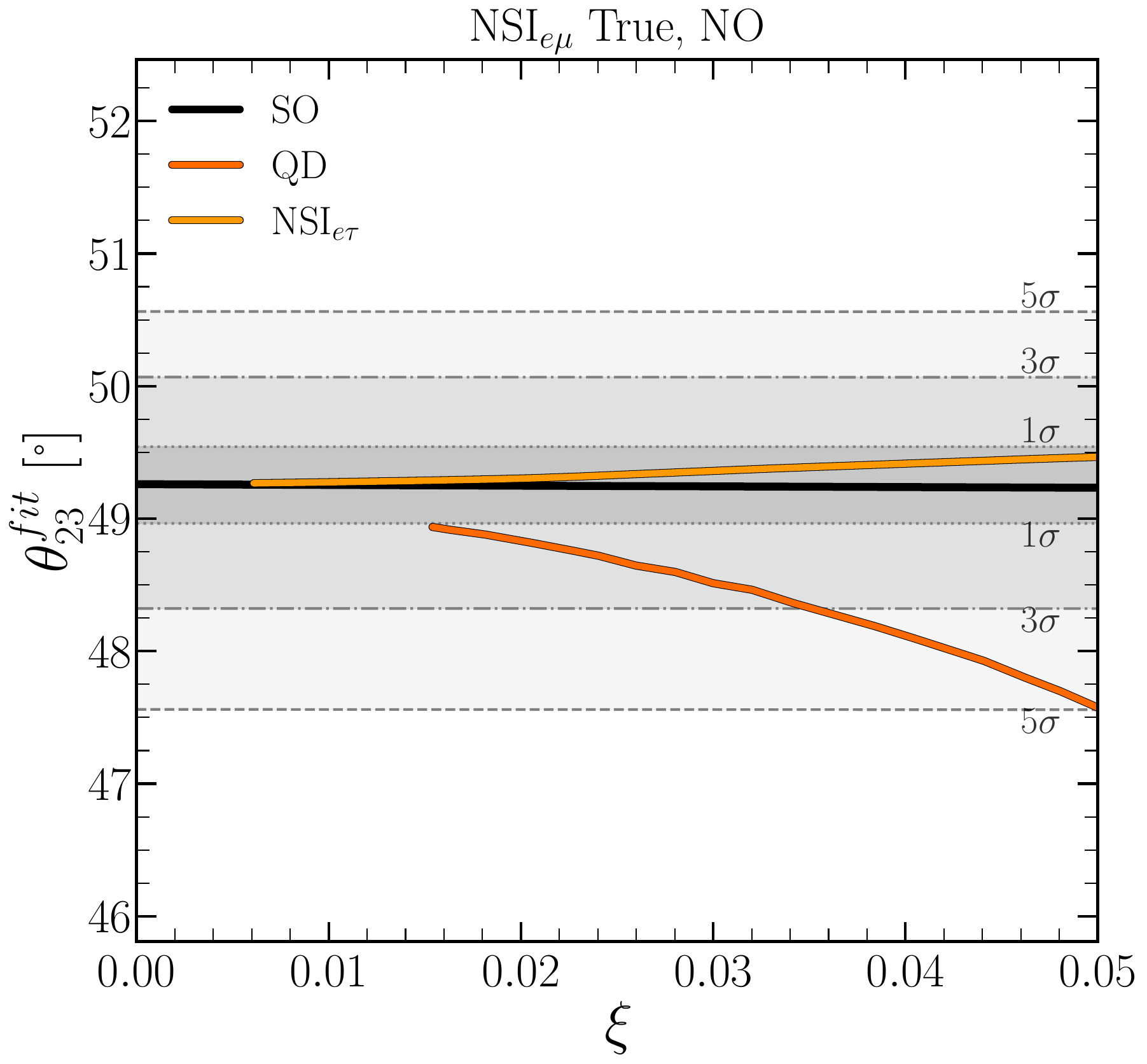}
	\end{subfigure}

        \vspace{-0.3cm}
	\caption{NSI$_{e\mu}$ as a True Model with $\delta_{CP}^{true} = -90^\circ$.}
 \label{A-Plots_NSI_EM_dcp_-90_th23}
\end{figure}

In Fig.~\ref{A-Plots_NSI_EM_dcp_-90} 
we 
present our results when 
the true simulated data is generated for
NSI$_{e\mu}$. In this case, only the plot of $\delta_{CP}^{fit}$ versus $\xi$ satisfies the filter criteria. For NO with $\delta_{CP}^{true} = -90^\circ (180^\circ)$, QD and $\text{NSI}_{e\tau}$ are the only BSO non-trivial solutions, while for IO, the non-trivial solutions are VEP and $\text{NSI}_{e\tau}$. The $\text{NSI}_{e\tau}$ is the sole non-trivial solution for IO with $\delta_{CP}^{true} = 180^\circ$. Within this context, the IO case shows slightly higher discrepancies between 
$\delta_{CP}^{fit}$ and $\delta_{CP}^{true}$ compared to the NO case. For NO, in general, the SO and QD are rather close to 
the 3$\sigma$ discrepancy at 
$\xi=0.05$ with $\delta_{CP}^{fit} \sim -117^\circ$ and $161^\circ$, for  $\delta_{CP}^{true}=-90^\circ$ and $180^\circ$, respectively. For IO and 
$\delta_{CP}^{true} = -90^\circ$, $\text{NSI}_{e\tau}$ comfortably exceeds the 3$\sigma$ discrepancy at 
$\xi=0.039$ with $\delta_{CP}^{fit} = -125.99^\circ$, while VEP is rather close at $\xi=0.05$ with $\delta_{CP}^{fit} = -120.30^\circ$. On the other hand, the SO scenario almost attains the 3$\sigma$ discrepancy 
for $\xi=0.05$ with $\delta_{CP}^{fit} = -163.71^\circ$, for IO and $\delta_{CP}^{true} = 180^\circ$. In Fig.~\ref{A-Plots_NSI_EM_dcp_-90_th23}, we present the plot of $\theta^{fit}_{23}$ versus $\xi$ for the NO scenario with $\delta_{CP}^{true} = -90^\circ$. In this case, only the QD scenario achieves a 3$\sigma$ separation at $\xi=0.035$, reaching nearly 5$\sigma$ at $\xi=0.05$.

\begin{figure}
	\centering
	\begin{subfigure}[t]{1.00\textwidth}
		\centering
		\includegraphics[width=\linewidth]{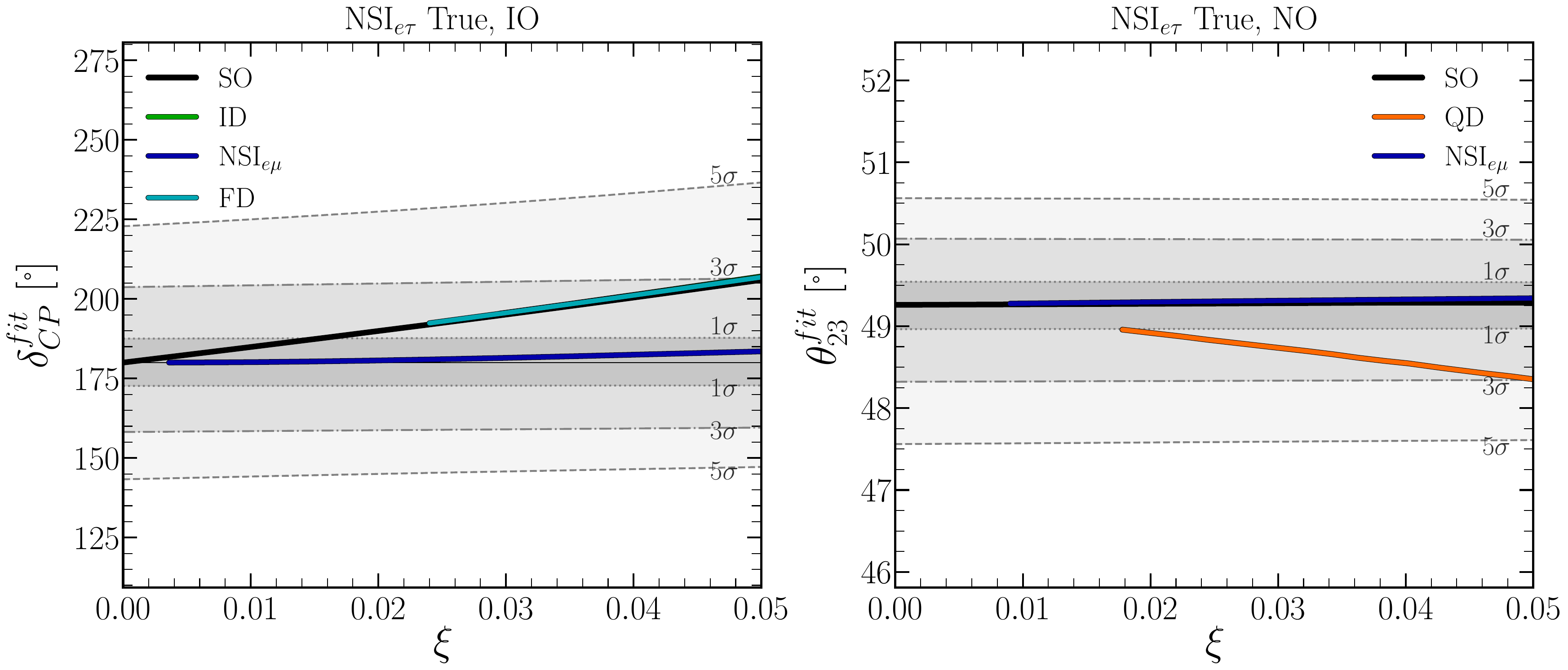}
	\end{subfigure}

        \vspace{-0.3cm}
	\caption{NSI$_{e\tau}$ as a True Model. On the left $\delta_{CP}^{true} = 180^\circ$ and on the right $\delta_{CP}^{true} = -90^\circ$. }
 \label{A-Plots_NSI_ET_dcp_-90_180_th23_dcp}
\end{figure}

In Fig.~\ref{A-Plots_NSI_ET_dcp_-90_180_th23_dcp}
two plots are displayed: 
$\delta_{CP}^{fit}$ versus $\xi$ 
for IO with $\delta_{CP}^{true} = 180^\circ$ and $\theta^{fit}_{23}$ versus $\xi$ for NO with $\delta_{CP}^{true} = -90^\circ$. 
For IO, the BSO trivial solutions are QD and VEP, in the case of NO, they are 
ID, FD, and VEP. 
For $\delta_{CP}^{fit}$, ID, FD and SO reach the 3$\sigma$ at $\xi=0.05$, for $\theta^{fit}_{23}$, only QD attains the 
3$\sigma$ at $\xi=0.05$.



\subsection{\textbf{ID}}

\begin{figure}
	\centering
	\begin{subfigure}[t]{1.0\textwidth}
		\centering
		\includegraphics[width=\linewidth]{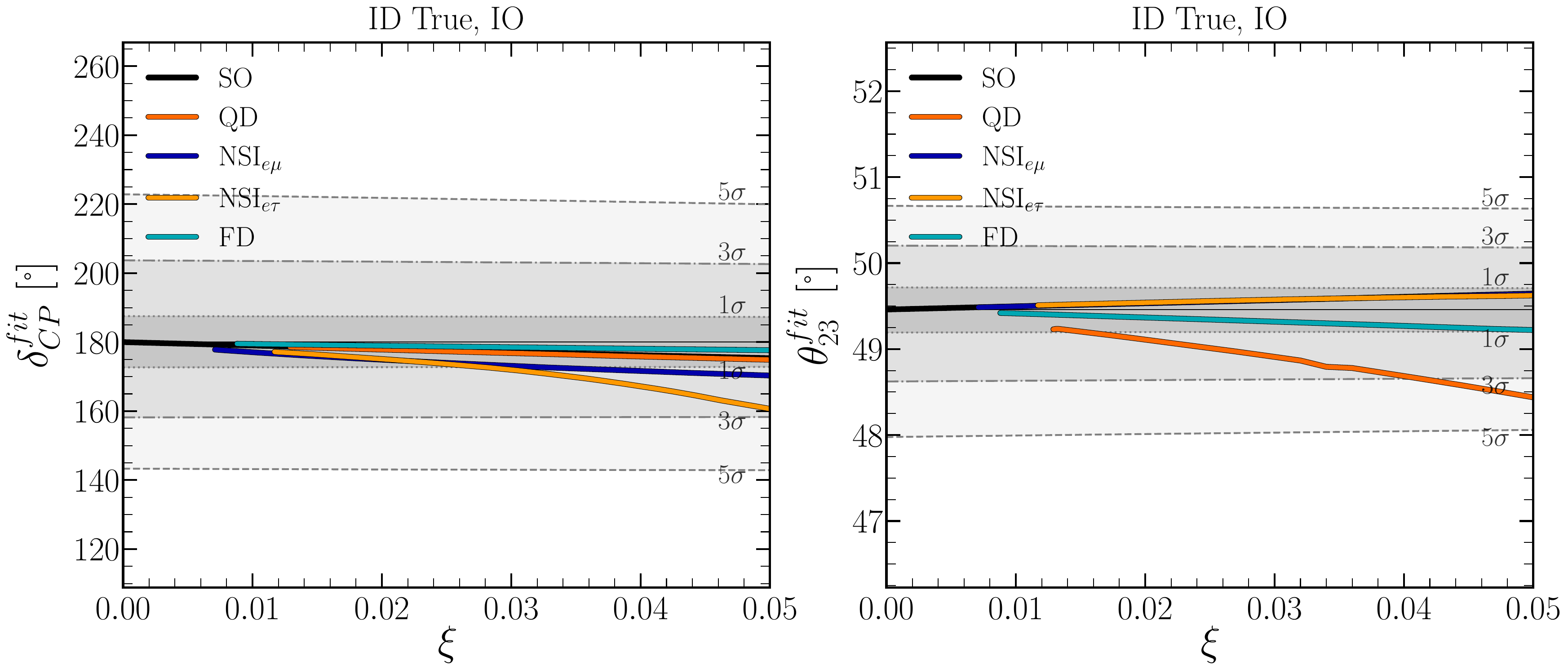}
	\end{subfigure}

        \vspace{-0.3cm}
	\caption{ID as a True Model with $\delta_{CP}^{true} = 180^\circ$.}
 \label{A-Plots_ID_180}
\end{figure}

In Fig.~\ref{A-Plots_ID_180}, we display the plots of 
$\delta_{CP}^{fit}$ vs $\xi$ 
and 
$\theta_{23}^{fit}$ vs $\xi$, both for IO and $\delta_{CP}^{true} =180^\circ$. VEP becomes the only trivial BSO solution when the ID is used as the true assumption. For $\delta_{CP}^{fit}$, NSI$_{e\tau}$ is the singular model that achieves a deviation approaching 3$\sigma$ at $\xi=0.05$. For $\theta_{23}^{fit}$, QD surpasses the 3$\sigma$ separation at $\xi=0.041$, while the other models remain below this threshold. 

\subsection{\textbf{FD}}
\begin{figure}
        \vspace{-0.8cm}
	\centering
	\begin{subfigure}[t]{1.00\textwidth}
		\centering
		\includegraphics[width=\linewidth]{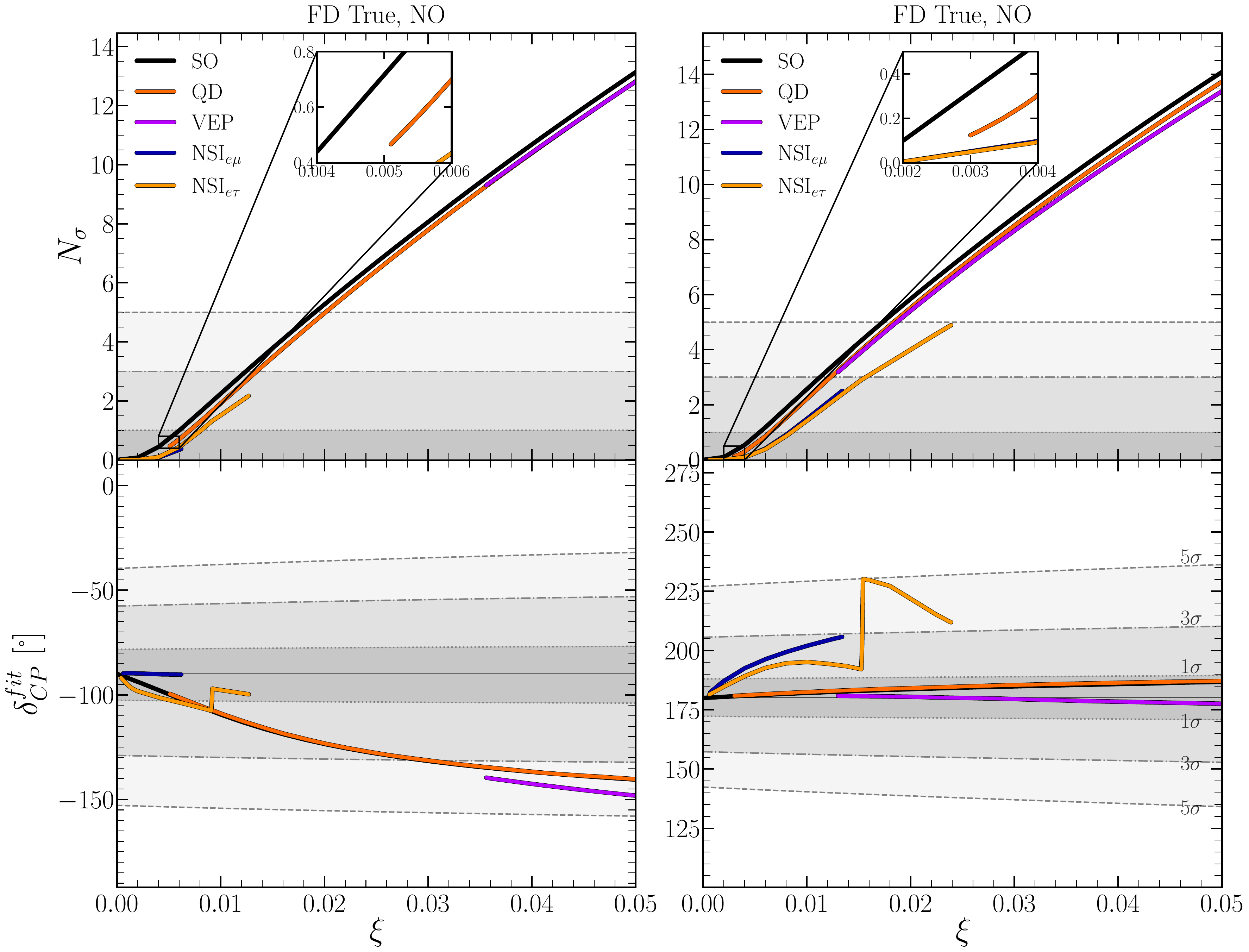}
	\end{subfigure}

        \vspace{-0.3cm}
	\caption{FD as a True Model. The left plots have $\delta_{CP}^{true} = -90^\circ$ and right plots has $\delta_{CP}^{true} = 180^\circ$.}
  \label{A-Plots_FD_Ns_dcp_-90_180_Ns_dcp}
\end{figure}

We show in Fig.~\ref{A-Plots_FD_Ns_dcp_-90_180_Ns_dcp} the plots of
$N_\sigma$ vs $\xi$ and 
$\delta_{CP}^{fit}$ vs $\xi$, for $\delta_{CP}^{true} = -90^\circ$ and $\delta_{CP}^{true} = 180^\circ$, all correspond to the NO scenario. The BSO trivial solution is the ID model valid for both values of $\delta_{CP}^{true}$. For $N_\sigma$ and both values of $\delta_{CP}^{true}$, SO, QD, and VEP significantly exceed the 5$\sigma$ deviation. Notably, VEP, for $\delta_{CP}^{true} = -90^\circ$, starts above the 5$\sigma$ threshold. The 5$\sigma$ separation is reached by SO and QD at $\xi=0.019(0.017)$ and $\xi=0.020(0.018)$ for $\delta_{CP}^{true} = -90^\circ (180^\circ)$, respectively. VEP achieves the 5$\sigma$ separation at $\xi=0.019$, for $\delta_{CP}^{true} = 180^\circ$, with NSI$_{e\tau}$ coming very close to this deviation. In the plot $\delta_{CP}^{fit}$ vs $\xi$ we have that, for $\delta_{CP}^{true} = -90^\circ$, SO and QD reach the 3$\sigma$ level at $\xi=0.029$ and $\xi=0.030$, with $\delta_{CP}^{fit}=-131.27^\circ$ and $-131.29^\circ$, respectively, whereas the VEP hypothesis starts above the 3$\sigma$ threshold. For $\delta_{CP}^{true} = 180^\circ$, the NSI$_{e\mu}$ hypothesis comes quite close to the 3$\sigma$ deviation at $\xi=0.013$, with $\delta_{CP}^{fit} = 205.77^\circ$. Meanwhile, NSI$_{e\tau}$ undergoes a steep change, reaching the 5$\sigma$ threshold at $\xi=0.015$ with $\delta_{CP}^{fit} = 230.09^\circ$.

\begin{figure}
        \vspace{-0.8cm}
	\centering
	\begin{subfigure}[t]{1.00\textwidth}
		\centering
		\includegraphics[width=\linewidth]{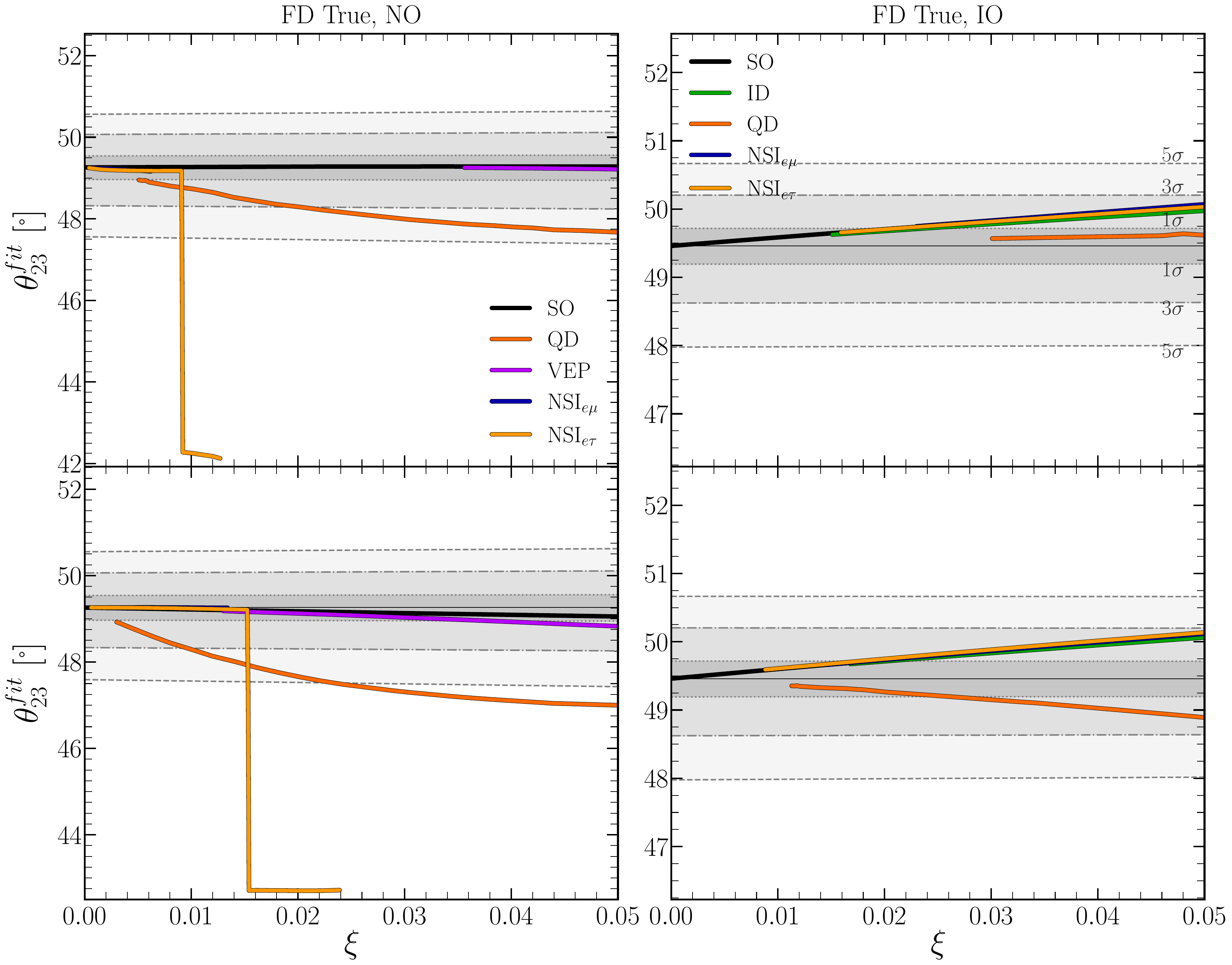}
	\end{subfigure}

        \vspace{-0.3cm}
	\caption{FD as a True Model. Plots at the top was made with $\delta_{CP}^{true} = -90^\circ$ and at the bottom was made with $\delta_{CP}^{true} = 180^\circ$.}
  \label{A-Plots_FD_dcp_-90_180_th23}
\end{figure}

In Fig.~\ref{A-Plots_FD_dcp_-90_180_th23}, we present the plots of $\theta^{fit}_{23}$ versus $\xi$ for NO and IO, corresponding to $\delta_{CP}^{true} = -90^\circ$ and $\delta_{CP}^{true} = 180^\circ$. For NO and $\delta_{CP}^{true} = -90^\circ$, the QD hypothesis crosses the 3$\sigma$ threshold at $\xi=0.020$, with $\theta^{fit}_{23} = 48.29^\circ$. For NSI$_{e\tau}$, there is a steep change at $\xi=0.009$, far exceeding the 5$\sigma$ deviation and resulting in $\theta^{fit}_{23} = 42.28^\circ$. This change in $\theta^{fit}_{23}$ is correlated with the one observed in $\delta_{CP}^{fit}$ in the corresponding plot. For NO and 
$\delta_{CP}^{true} = 180^\circ$, the behavior is similar to before, with the QD hypothesis crossing the 3$\sigma$ threshold at $\xi=0.010$, and even crossing the 5$\sigma$
at $\xi=0.023$. The NSI$_{e\tau}$ 
suffers a steep change at $\xi=0.015$, far exceeding the 5$\sigma$ deviation and resulting in $\theta^{fit}_{23} = 42.71^\circ$, coinciding with the shift seen in
$\delta_{CP}^{fit}$. For IO and $\delta_{CP}^{true} = -90^\circ$, the NSI$_{e\mu}$ hypothesis is the closest to the 3$\sigma$ level at $\xi=0.05$, with $\theta^{fit}_{23} = 50.07^\circ$. For $\delta_{CP}^{true} = 180^\circ$, the hypotheses nearest to the 3$\sigma$ deviation are NSI$_{e\tau}$, SO, and NSI$_{e\mu}$ at $\xi=0.05$, with $\theta^{fit}_{23} \sim 50.12^\circ$.

\begin{figure}
        \vspace{-0.8cm}
	\centering
	\begin{subfigure}[t]{1.00\textwidth}
		\centering
		\includegraphics[width=\linewidth]{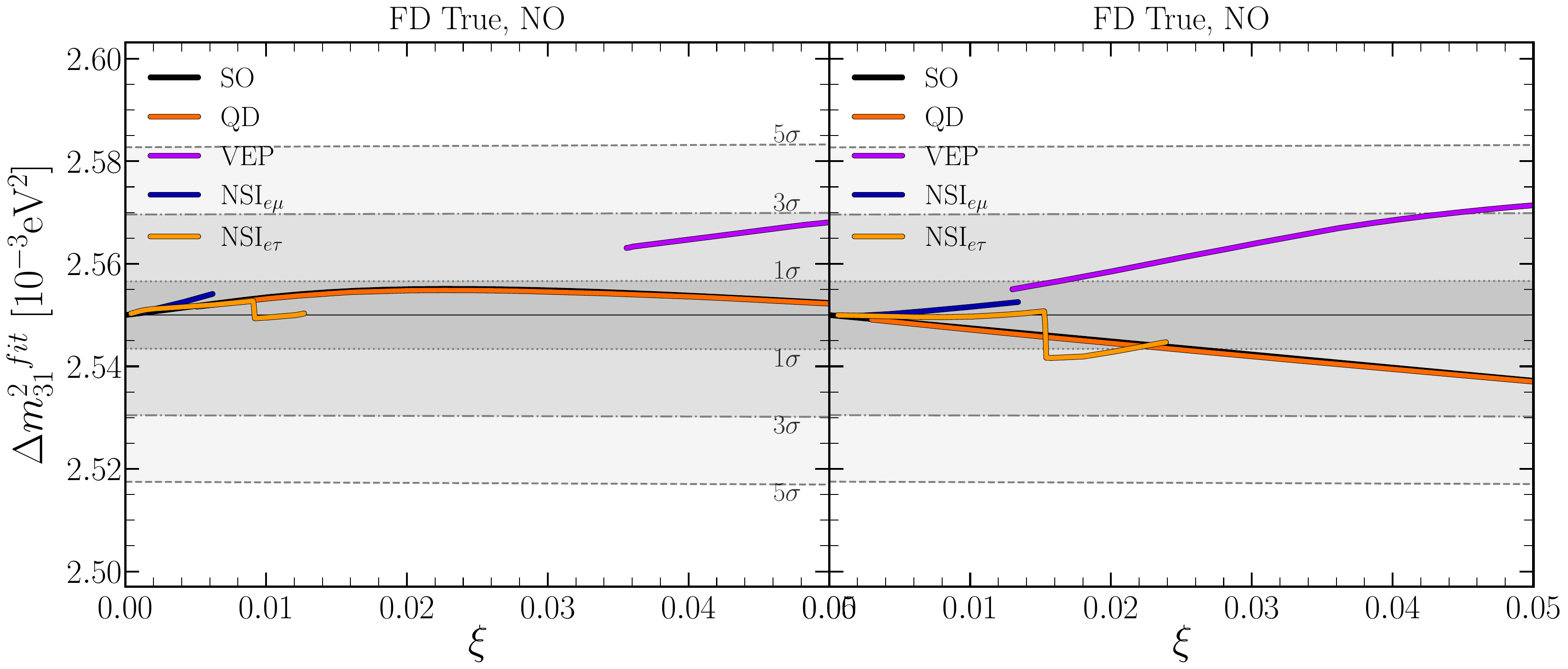}
	\end{subfigure}

        \vspace{-0.3cm}
	\caption{FD as a True Model. Left plot has $\delta_{CP}^{true} = -90^\circ$ and right plot has $\delta_{CP}^{true} = 180^\circ$.}
  \label{A-Plots_FD_dcp_-90_180_dm31}
\end{figure}

In Fig.~\ref{A-Plots_FD_dcp_-90_180_dm31}, we display the plots of 
$\Delta m^2_{31}{}^{fit}$ versus 
$\xi$ for NO, corresponding to 
$\delta_{CP}^{true} = -90^\circ$ and 
$\delta_{CP}^{true} = 180^\circ$. VEP is the only model with a separation close to (or exceeding) the 3$\sigma$ deviation for $\delta_{CP}^{true} = -90^\circ (180^\circ)$ at $\xi=0.05 (0.044)$, with 
$\Delta m^2_{31}{}^{fit} = 2.568 (2.570) \times 10^{-3} \text{eV}^2$.


\subsection{\textbf{QD}}

\begin{figure}
        \vspace{-0.45cm}
	\centering
	\begin{subfigure}[t]{1.0\textwidth}
		\centering
		\includegraphics[width=\linewidth]{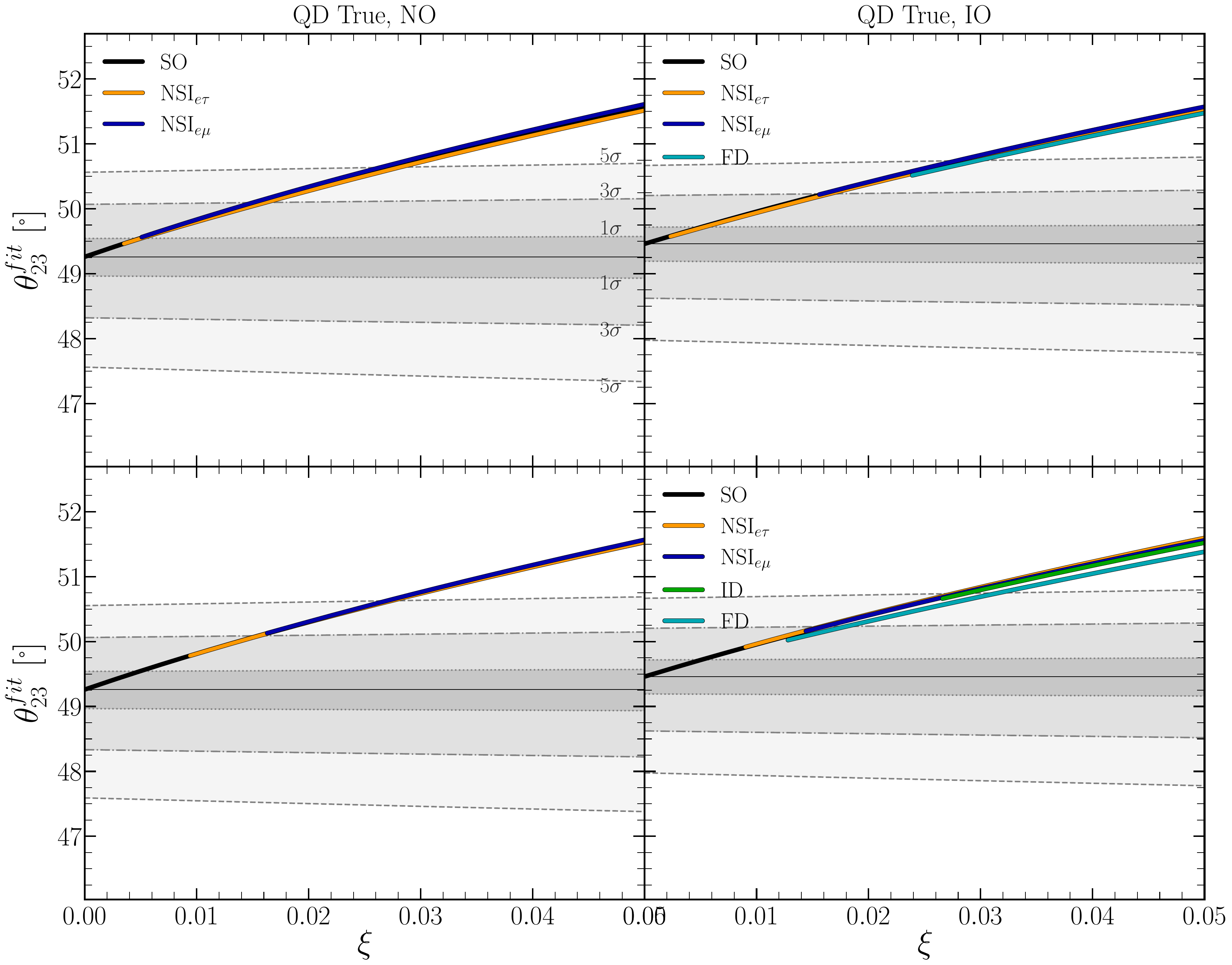}
	\end{subfigure}

        \vspace{-0.3cm}
	\caption{QD as a True Model. In the first row $\delta_{CP}^{true} = -90^\circ$ and in the second row $\delta_{CP}^{true} = 180^\circ$.}
  \label{A-Plots_QD_dcp_-90_180_th23}
\end{figure}

For QD, only the plots of $\theta^{fit}_{23}$ versus $\xi$ for both NO and IO, with $\delta_{CP}^{true} = -90^\circ$ and $\delta_{CP}^{true} = 180^\circ$, displayed in Fig.~\ref{A-Plots_QD_dcp_-90_180_th23}, satisfied our criteria. The non-trivial BSO solutions for NO are SO, NSI$_{e\mu}$, and NSI$_{e\tau}$, for both values of $\delta_{CP}^{true}$. For IO, the non-trivial solutions are SO, NSI$_{e\mu}$, NSI$_{e\tau}$, and FD for $\delta_{CP}^{true} = -90^\circ$; for $\delta_{CP}^{true} = 180^\circ$, the same solutions hold with the addition of ID. For NO and $\delta_{CP}^{true} = -90^\circ$, all the non-trivial solutions cross the 5$\sigma$ threshold, with SO, NSI$_{e\mu}$ and NSI$_{e\tau}$ surpassing the 3$\sigma$ level at $\xi = 0.015, 0.015$ and $0.016$, corresponding to $\theta^{fit}_{23} = 50.09^\circ, 50.09^\circ$ and $50.10^\circ$, respectively. For $\delta_{CP}^{true} = 180^\circ$, again, all the 
non-trivial solutions cross the 5$\sigma$ level, with SO and NSI$_{e\tau}$ surpassing the 3$\sigma$ level at $\xi = 0.015$ and $0.016$, corresponding to $\theta^{fit}_{23} = 50.09^\circ$ and $50.09^\circ$, respectively. The NSI$_{e\mu}$ hypothesis begins just above the 3$\sigma$ threshold. For IO and $\delta_{CP}^{true} = -90^\circ$, all the BSO non-trivial solutions cross the 5$\sigma$ level, with SO, NSI$_{e\mu}$, and NSI$_{e\tau}$ crossing the 3$\sigma$ threshold at $\xi = 0.016$, corresponding to $\theta^{fit}_{23} = 50.23^\circ$. The FD hypothesis emerges above the 3$\sigma$ level. For $\delta_{CP}^{true} = 180^\circ$, all the non-trivial solutions cross the 5$\sigma$ level, with SO, NSI$_{e\mu}$, NSI$_{e\tau}$, and FD crossing the 3$\sigma$ threshold at $\xi = 0.016$ (for SO, NSI$_{e\mu}$, and NSI$_{e\tau}$) and $\xi = 0.018$ (for FD), corresponding to $\theta^{fit}_{23} = 50.23^\circ$ in all cases. In this case, the ID hypothesis emerges above the 3$\sigma$ level.

\begin{table}[h]

  \vspace{-0.8cm}
  \centering
  \begin{tabular}{cc}
    \includegraphics[width=0.485\textwidth]{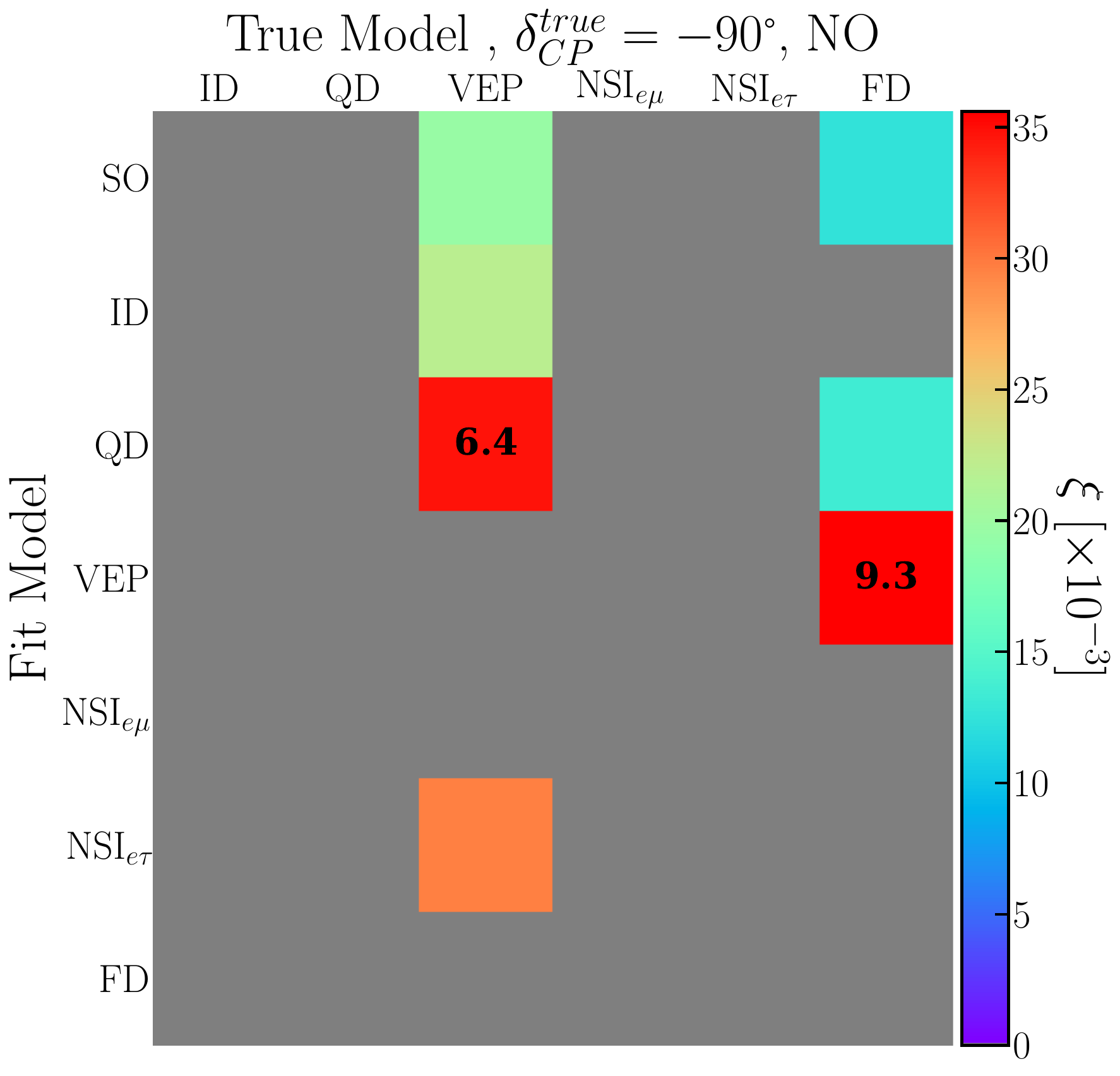} & \includegraphics[width=0.495\textwidth]{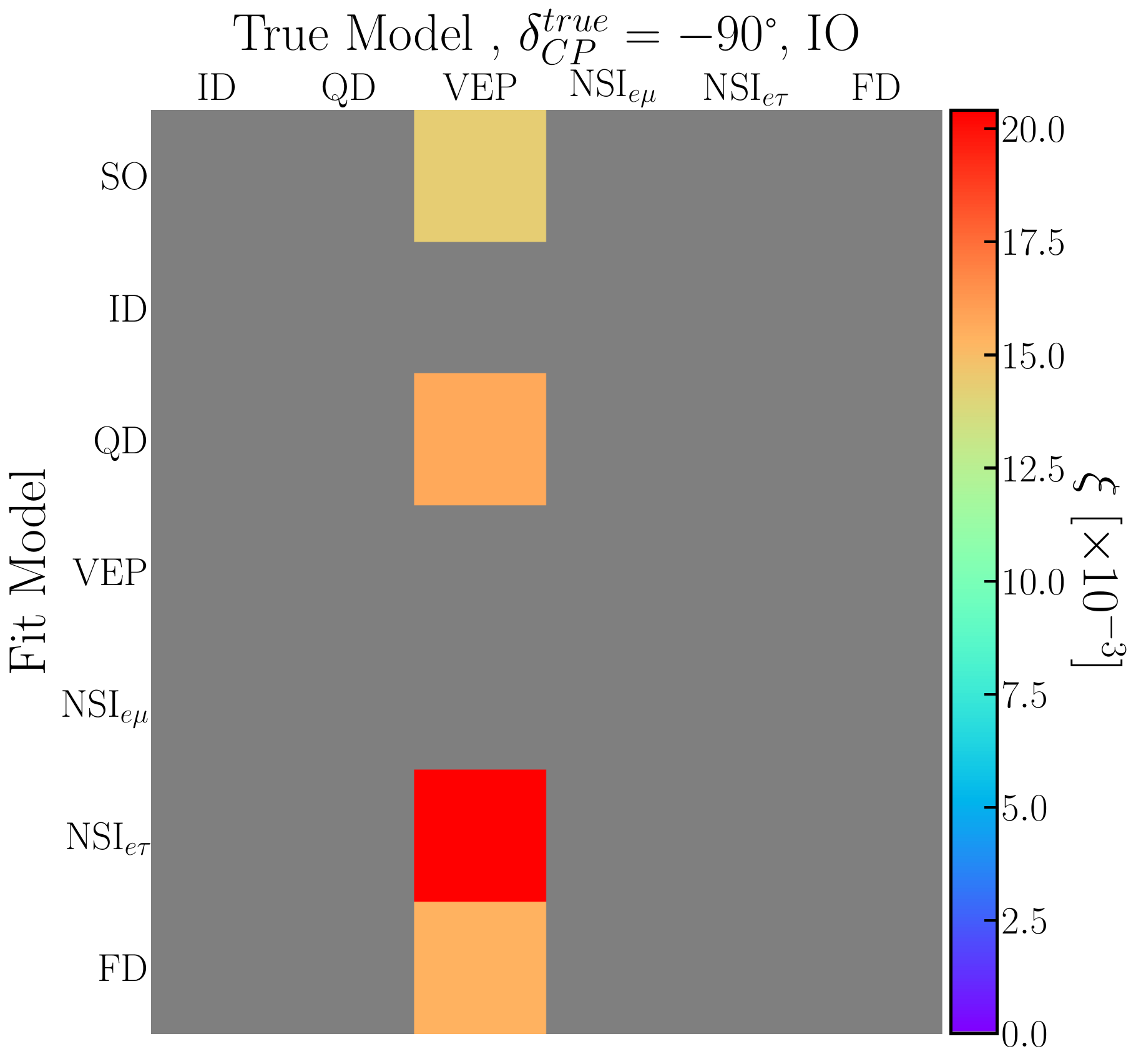} \\
    \includegraphics[width=0.49\textwidth]{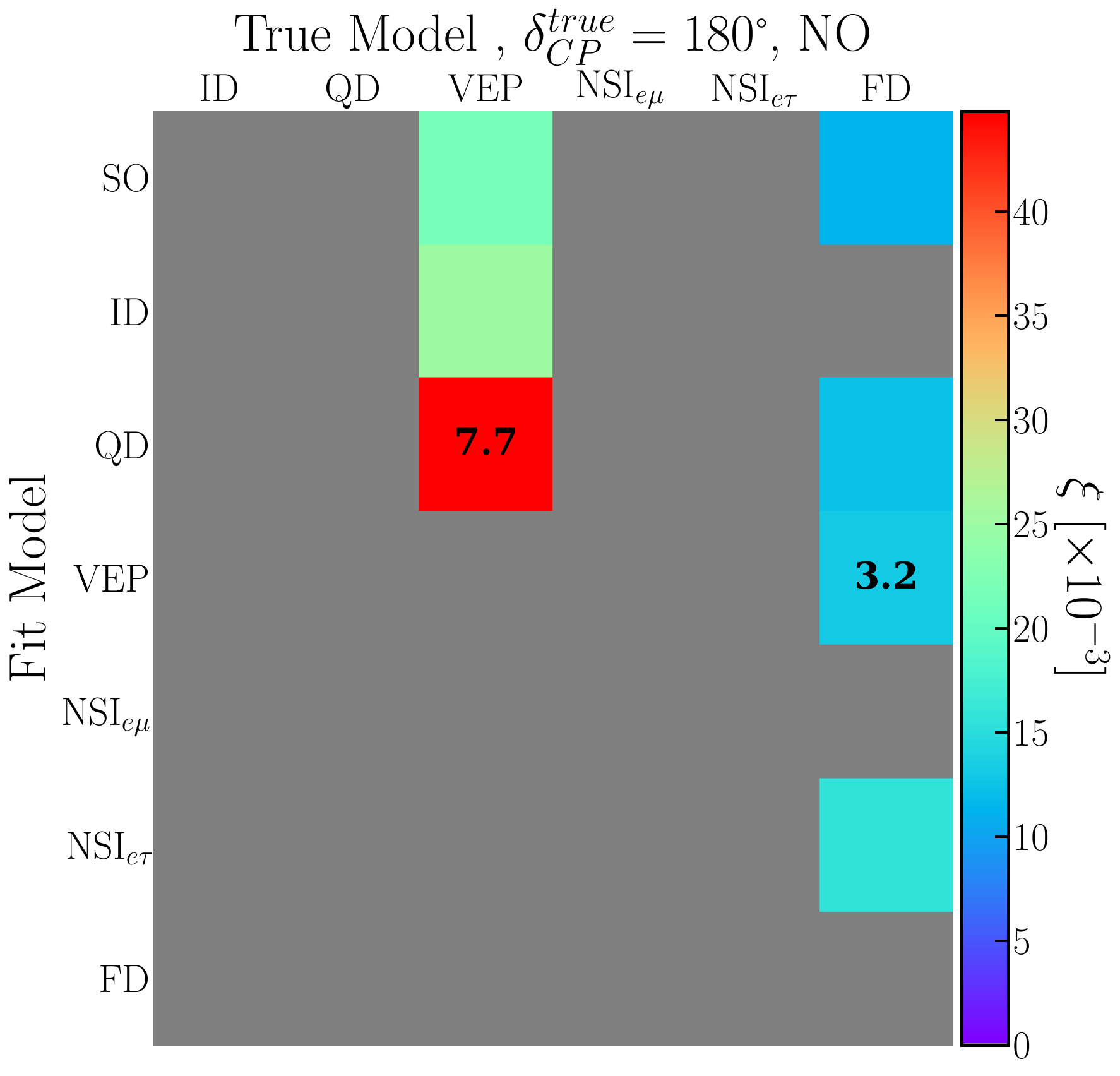} & 
    \includegraphics[width=0.49\textwidth]{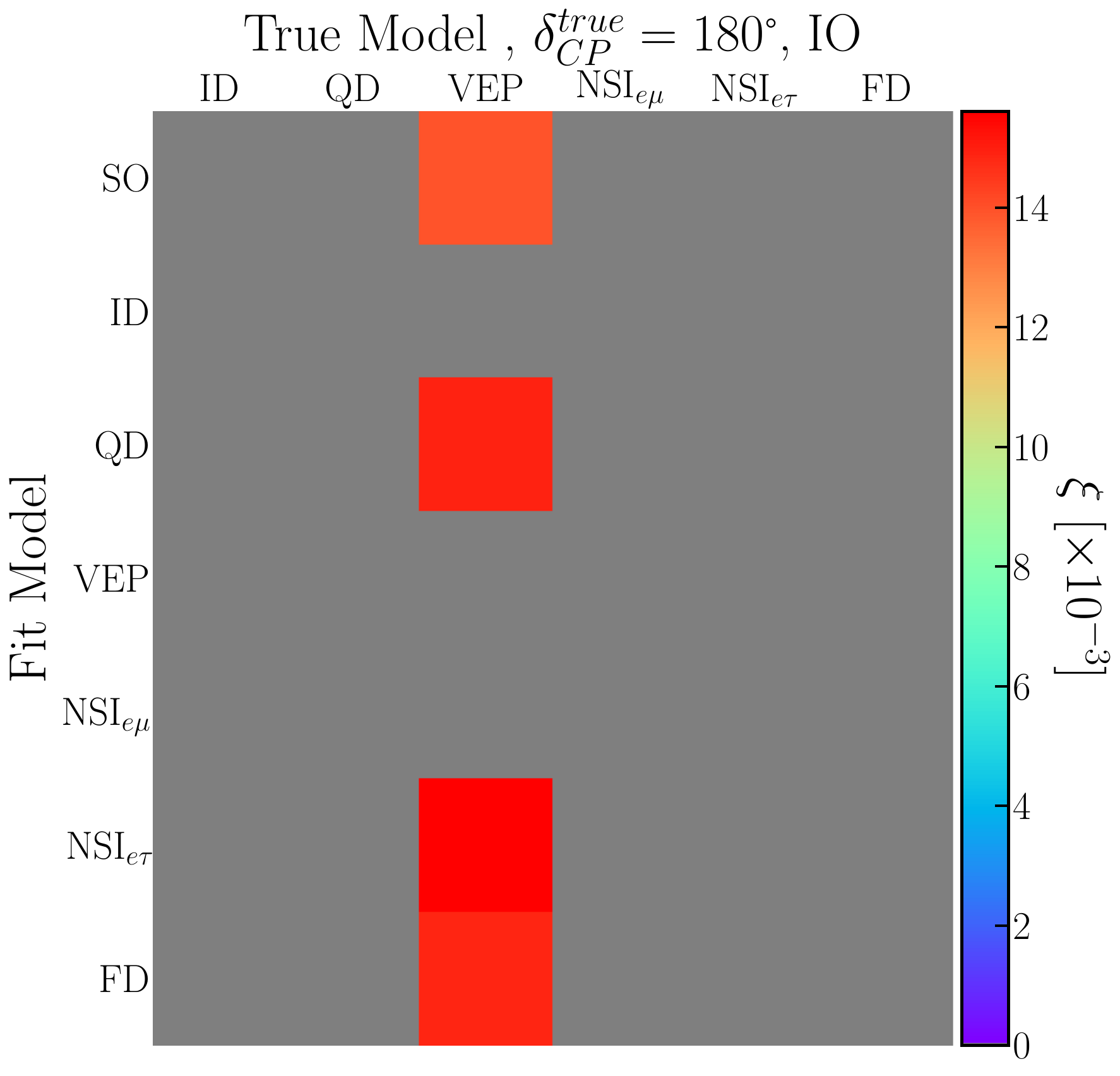} \\
  \end{tabular}
  \vspace{-0.4cm}
	\caption{Table of True Models vs Fit Models in which the value of $\xi$ is shown for which $N_\sigma = 3\sigma$ is obtained. The boxes displaying numbers indicate the cases where $N_\sigma > 3\sigma$ and the number shown is the $N_{\sigma}$ obtained.}
 \vspace{-0.5cm}
 \label{A-Cuadro_3_sigma_dcp_-90_180}
\end{table}

\begin{table}[h]
  \vspace{-0.8cm}
  \centering
  \begin{tabular}{cc}
    \includegraphics[width=0.492\textwidth]{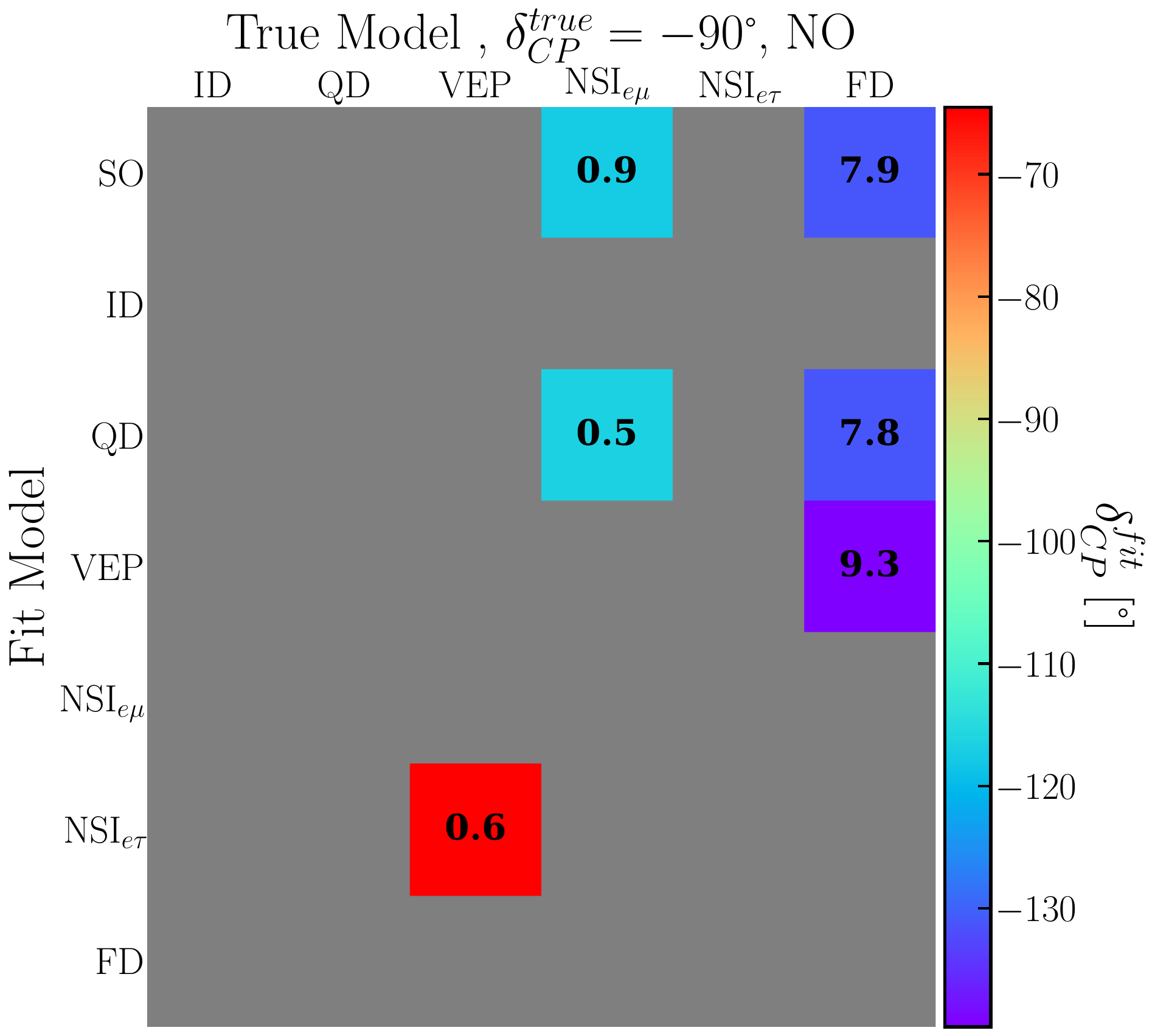} & \includegraphics[width=0.488\textwidth]{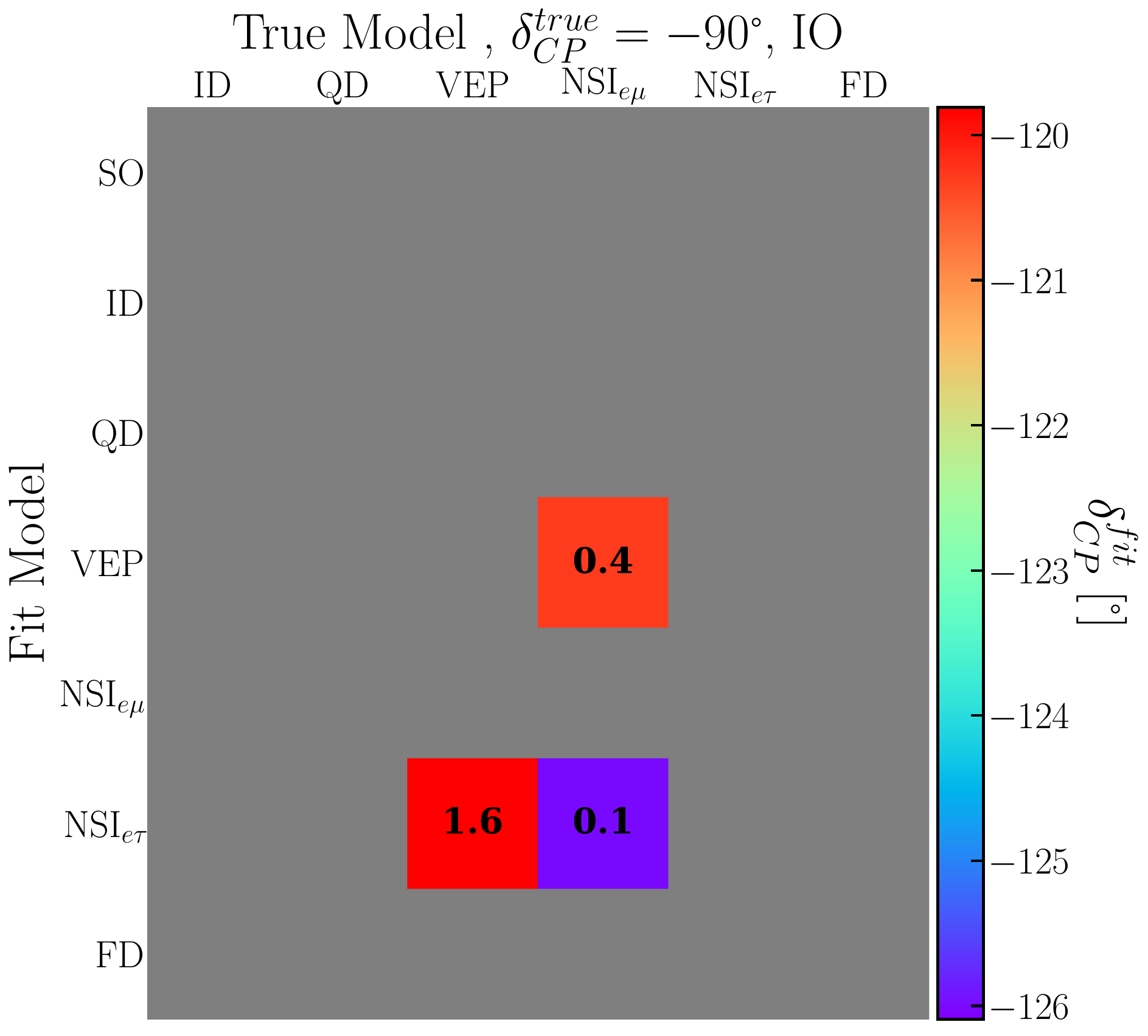} \\
    \includegraphics[width=0.49\textwidth]{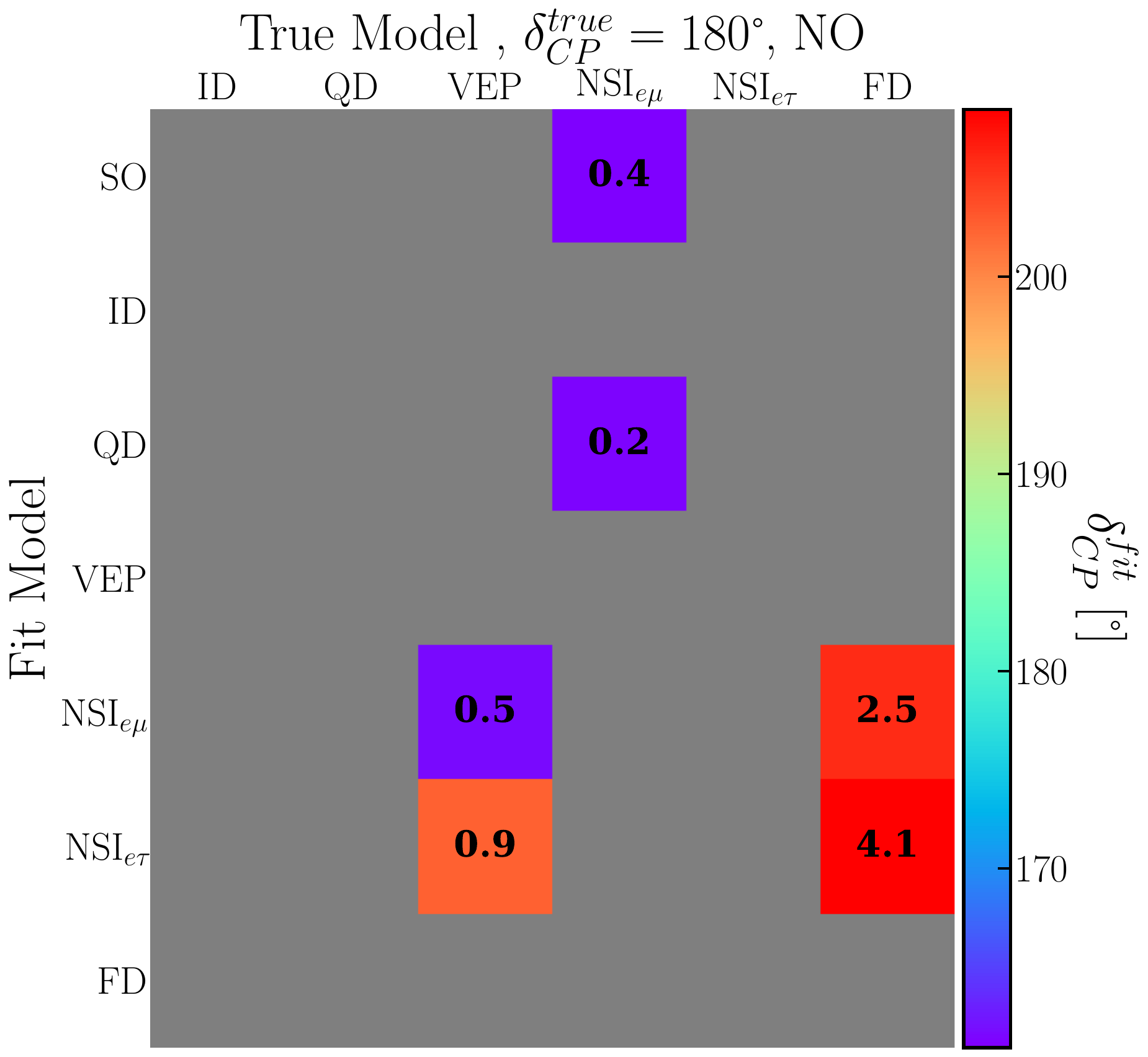} & \includegraphics[width=0.49\textwidth]{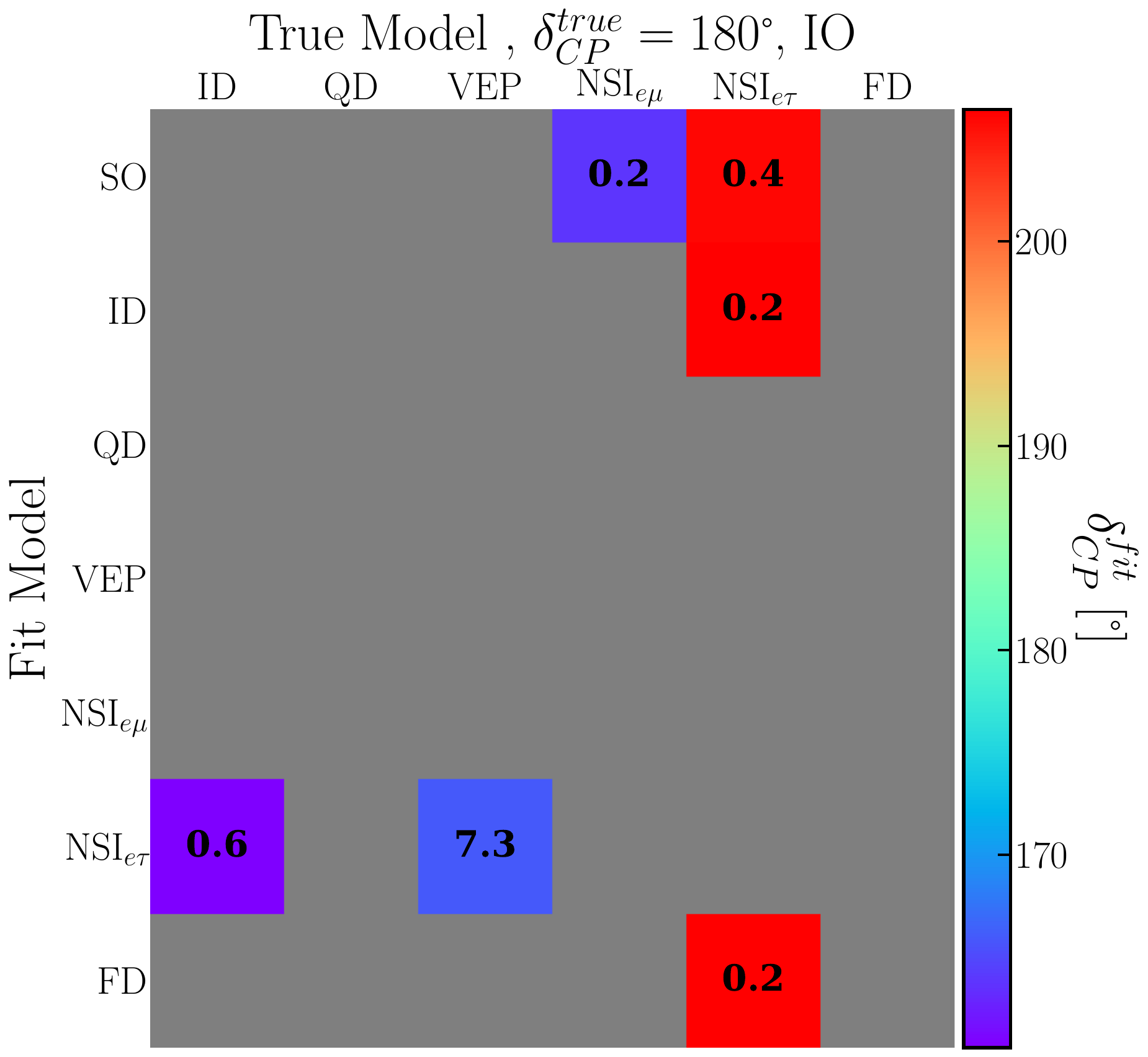} \\
  \end{tabular}
  \vspace{-0.3cm}
	\caption{Table of True Models vs Fit Models in which the values of $\delta^{fit}_{CP}$ are when the deviation respect to $\delta^{true}_{CP}$ is around $2.4 \sigma$. The boxes displaying numbers indicate the respective $N_\sigma$.}
 \label{A-Cuadro_dcp_3_sigma_dcp_-90_180}
\end{table}

\begin{table}[h]
  \vspace{-0.8cm}
  \centering
  \begin{tabular}{cc}
    \includegraphics[width=0.49\textwidth]{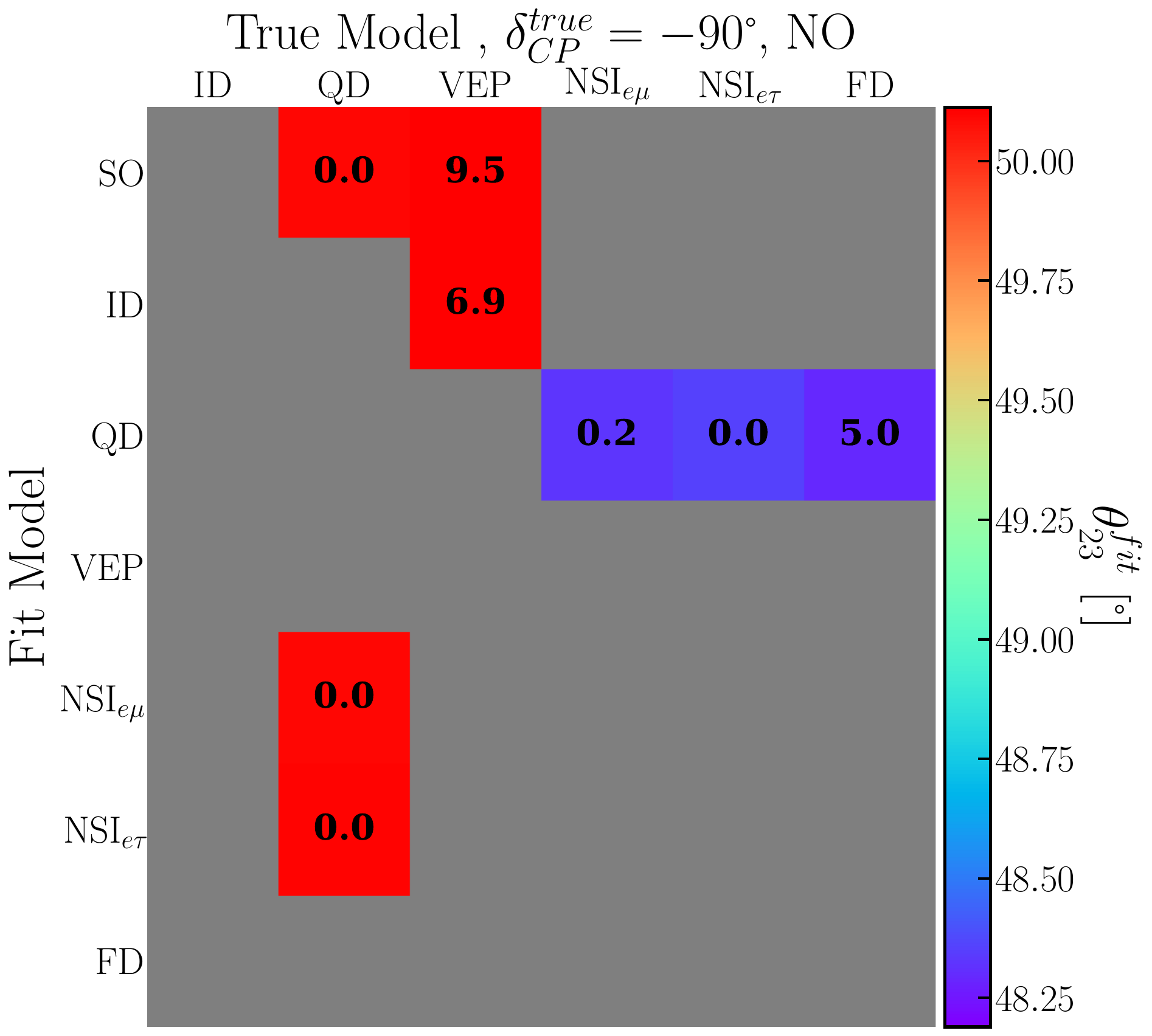} & \includegraphics[width=0.49\textwidth]{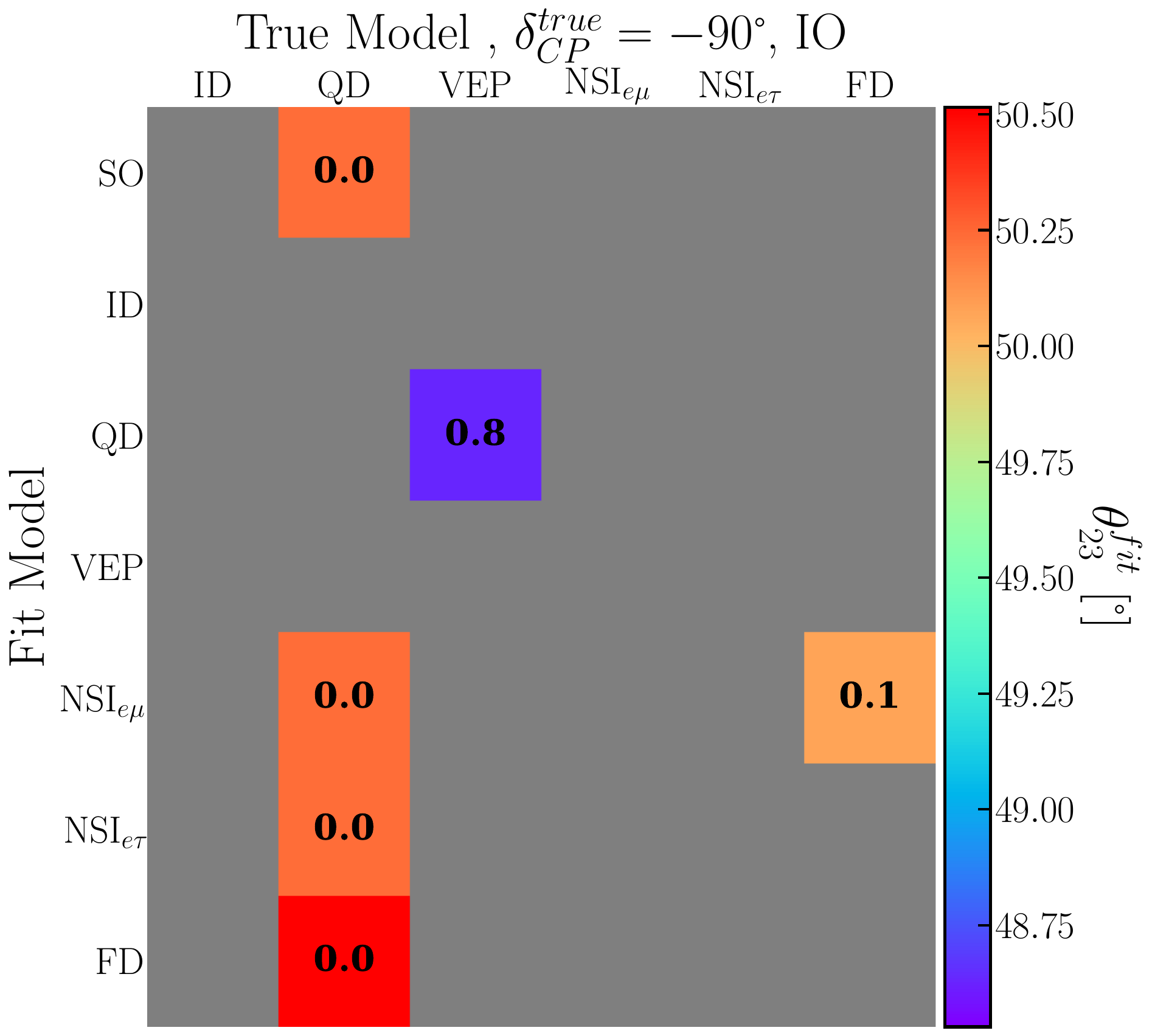} \\
    \includegraphics[width=0.49\textwidth]{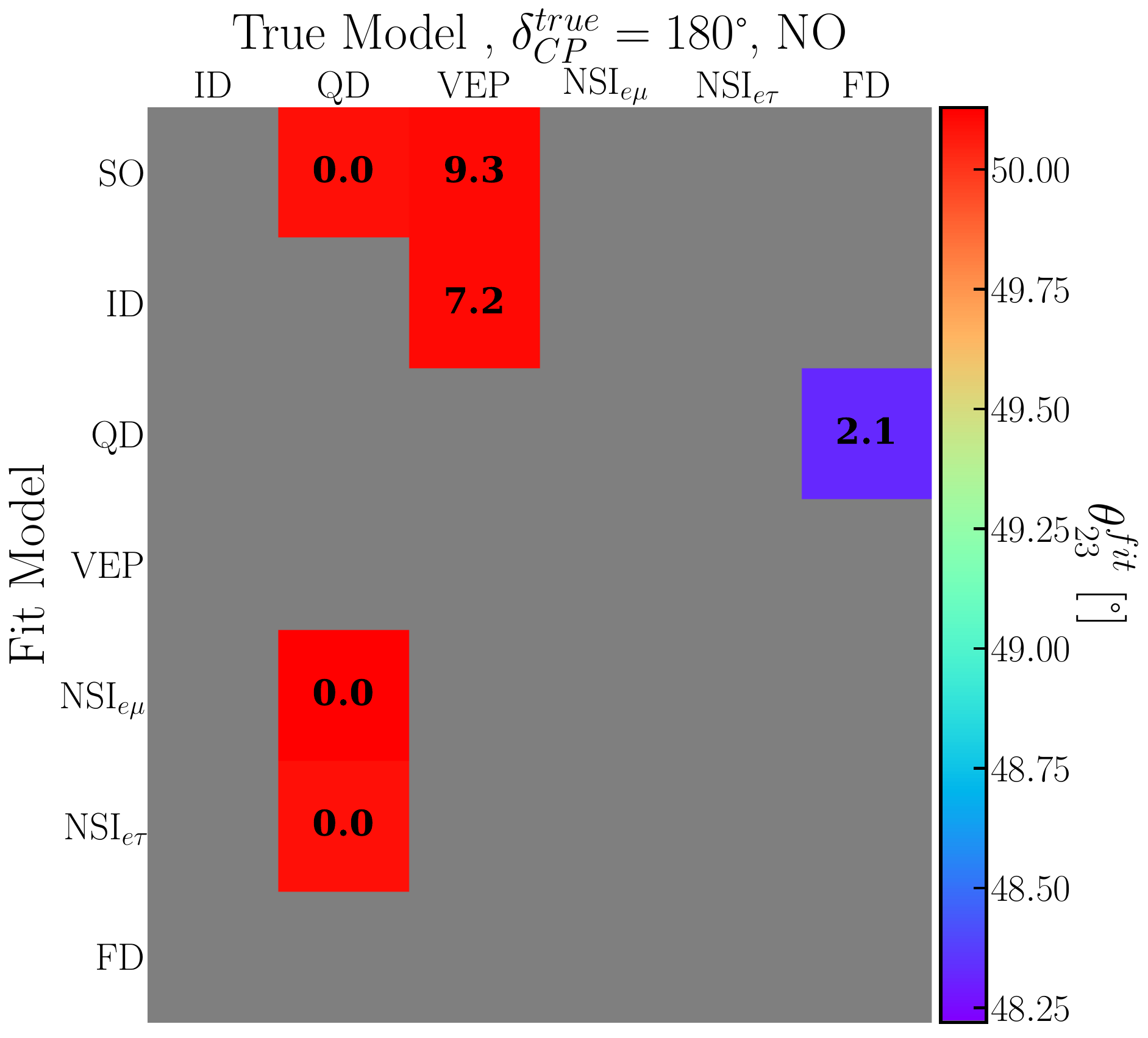} & \includegraphics[width=0.49\textwidth]{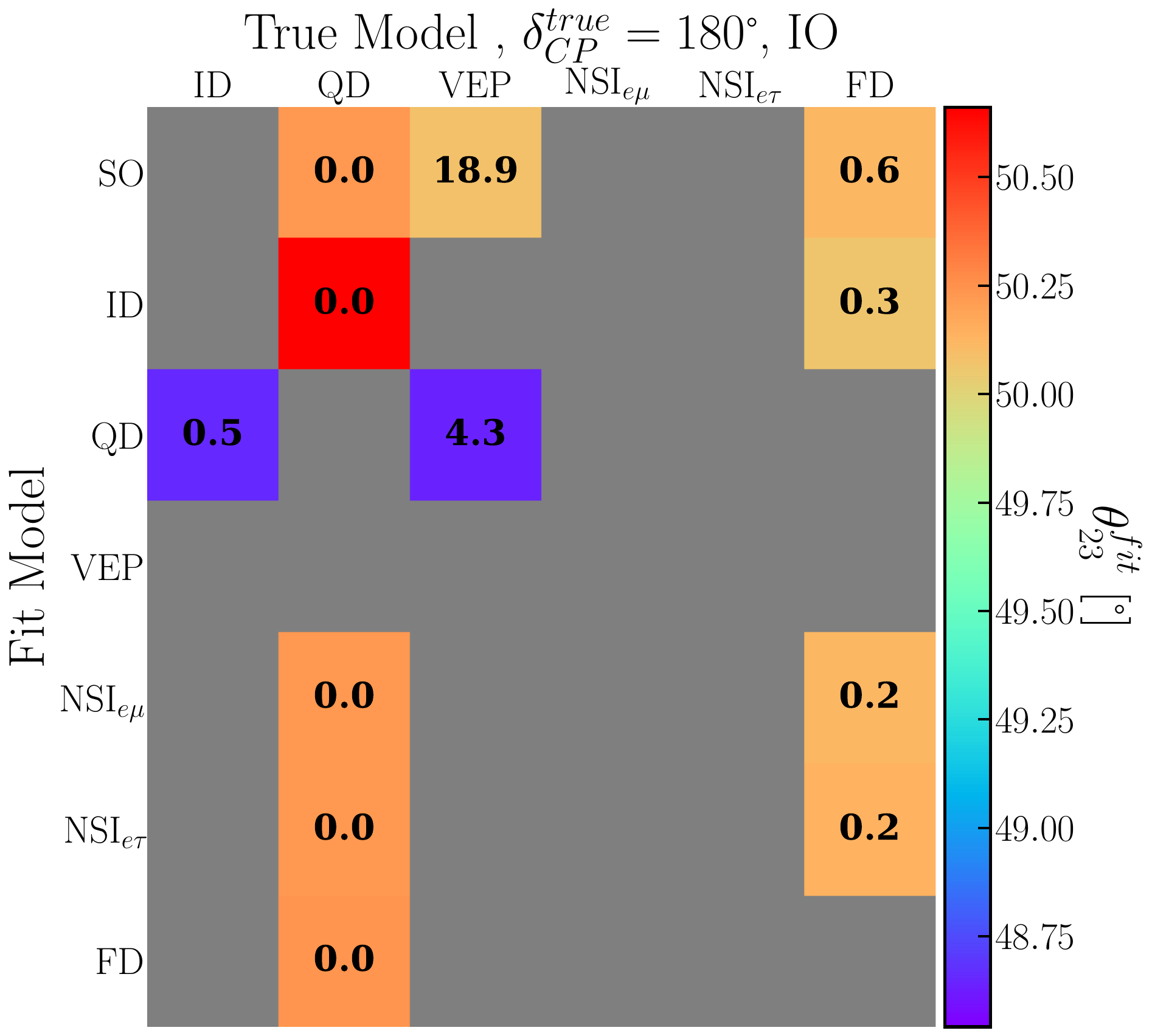} \\
  \end{tabular}
  \vspace{-0.3cm}
	\caption{Table of True Models vs Fit Models in which the values of $\theta_{23}$ are when the deviation respect to $\theta_{23}^{true}$ is around $2.4 \sigma$. The boxes displaying numbers indicate the respective $N_\sigma$.}
 \label{A-Cuadro_th23_3_sigma_dcp_-90_180}
\end{table}

\begin{table}[h]
  \vspace{-0.8cm}
  \centering
  \begin{tabular}{cc}
    \includegraphics[width=0.48\textwidth]{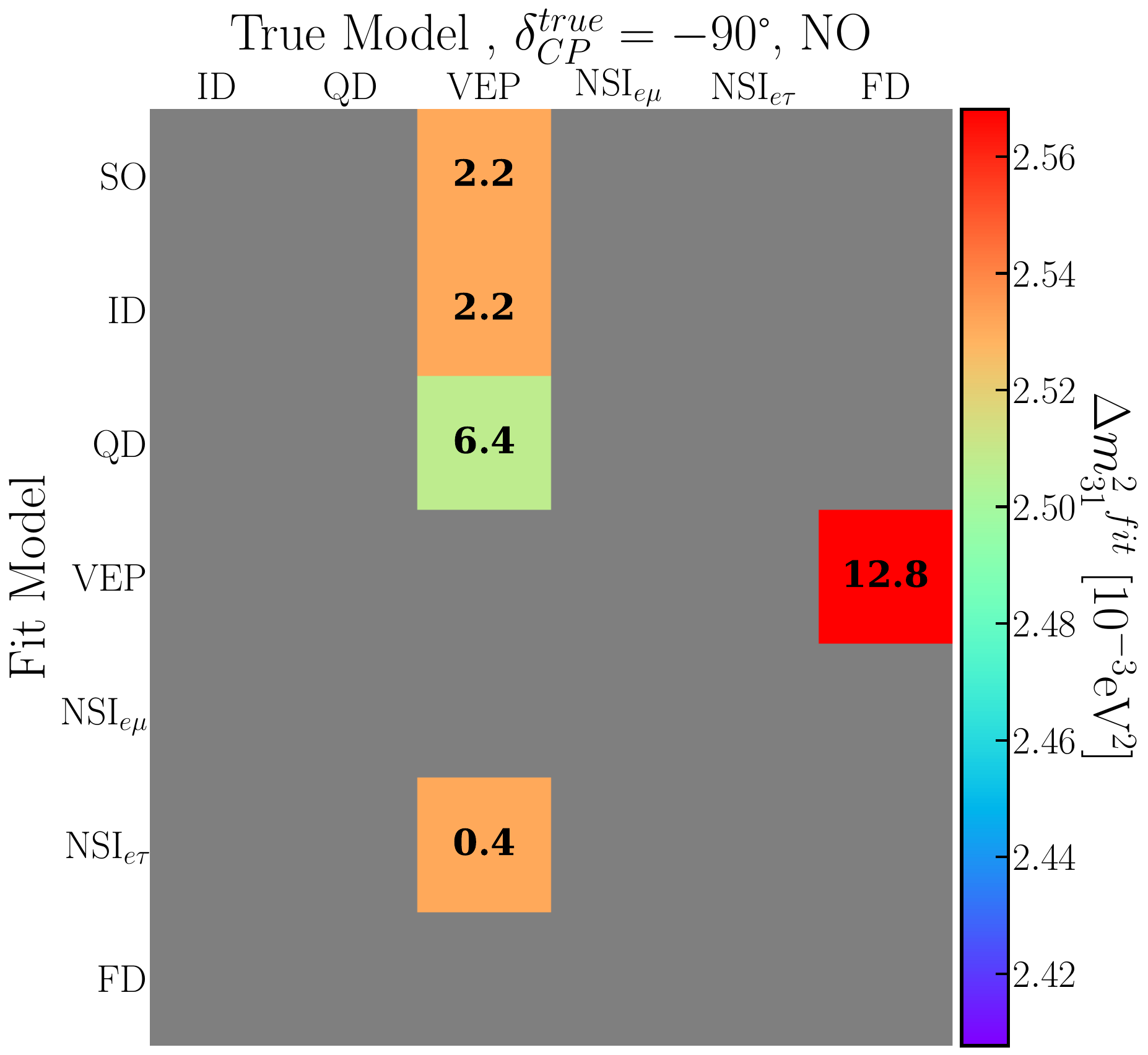} & 
    \includegraphics[width=0.48\textwidth]{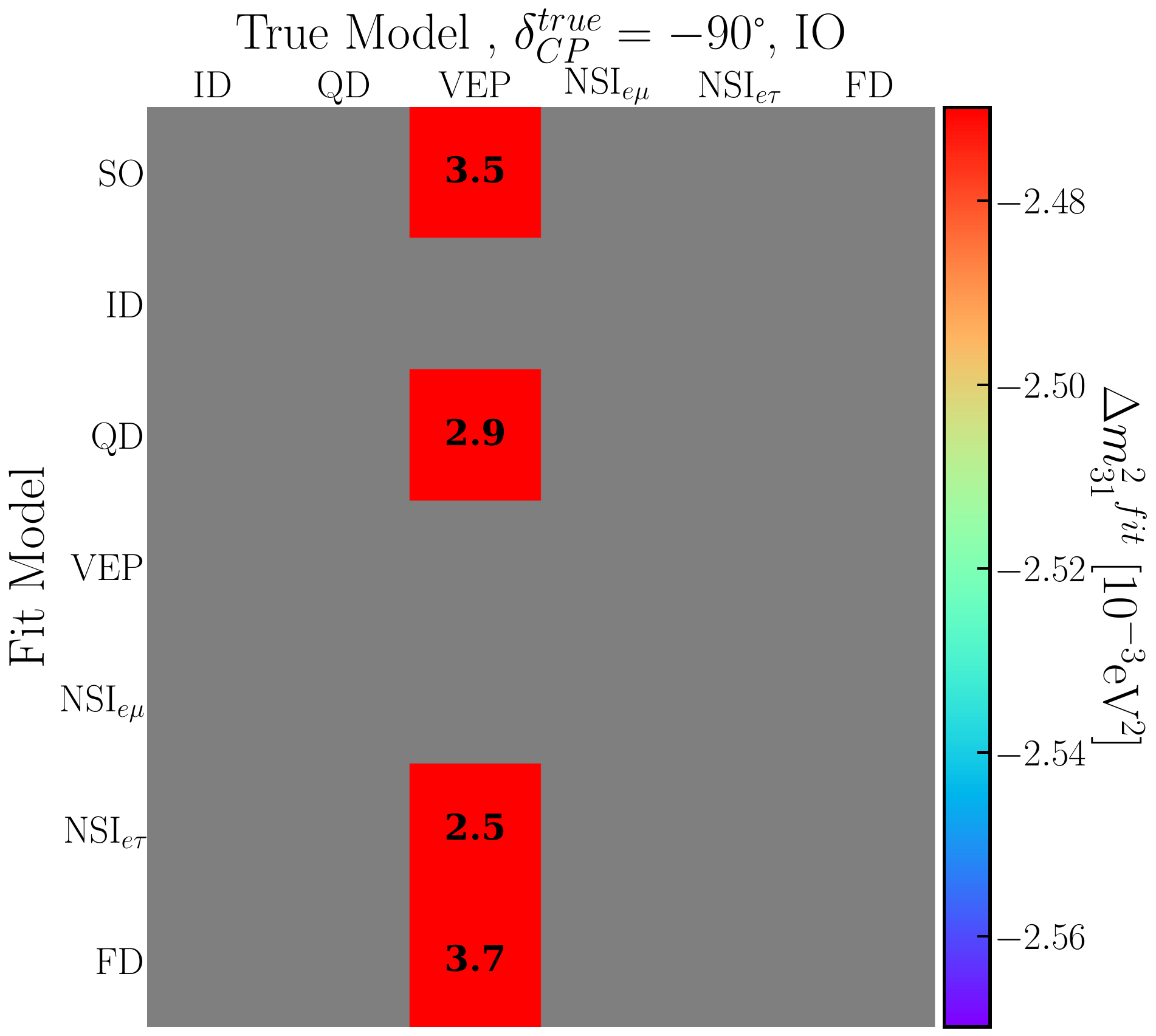} \\
    \includegraphics[width=0.48\textwidth]{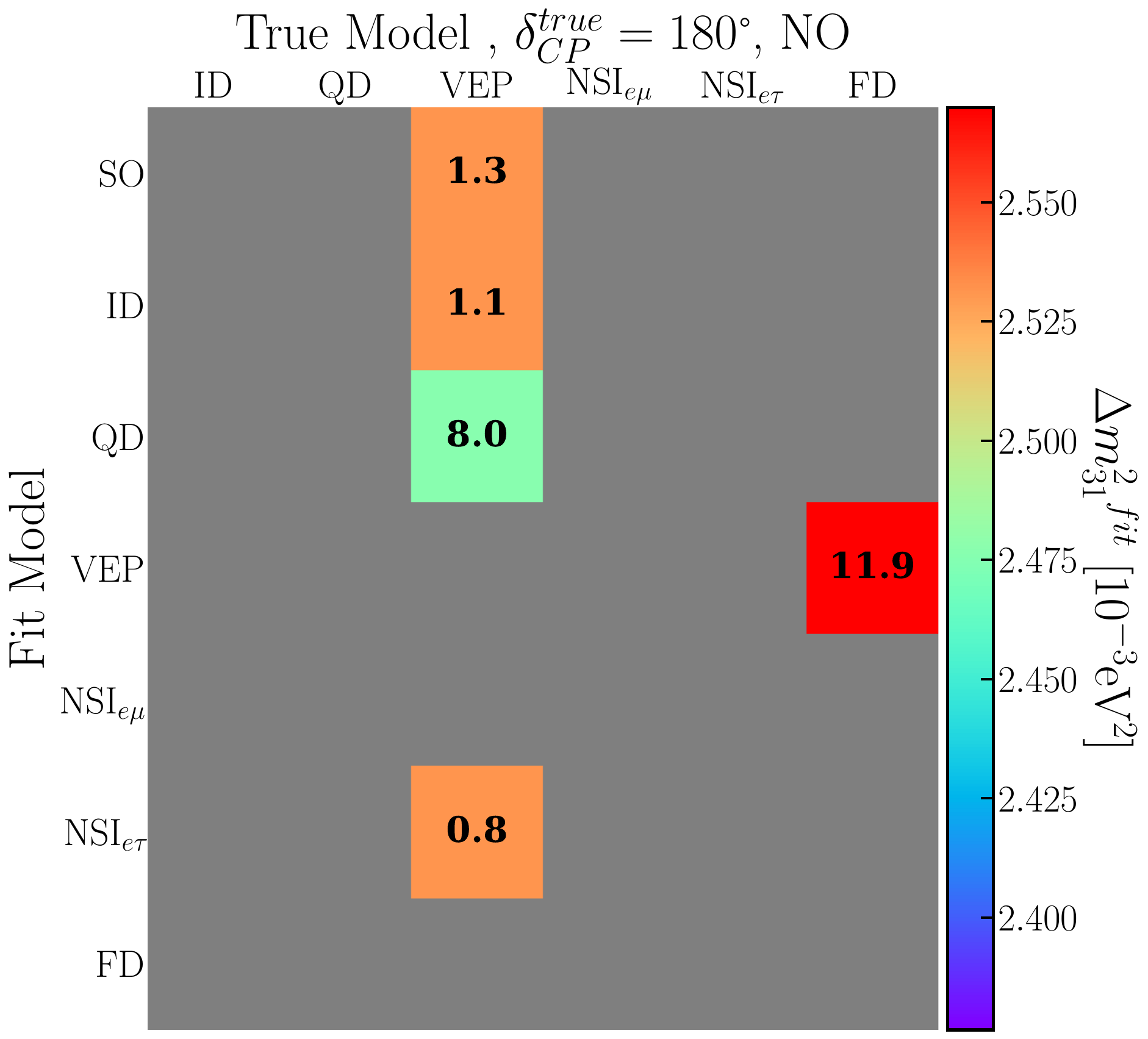} & 
    \includegraphics[width=0.48\textwidth]{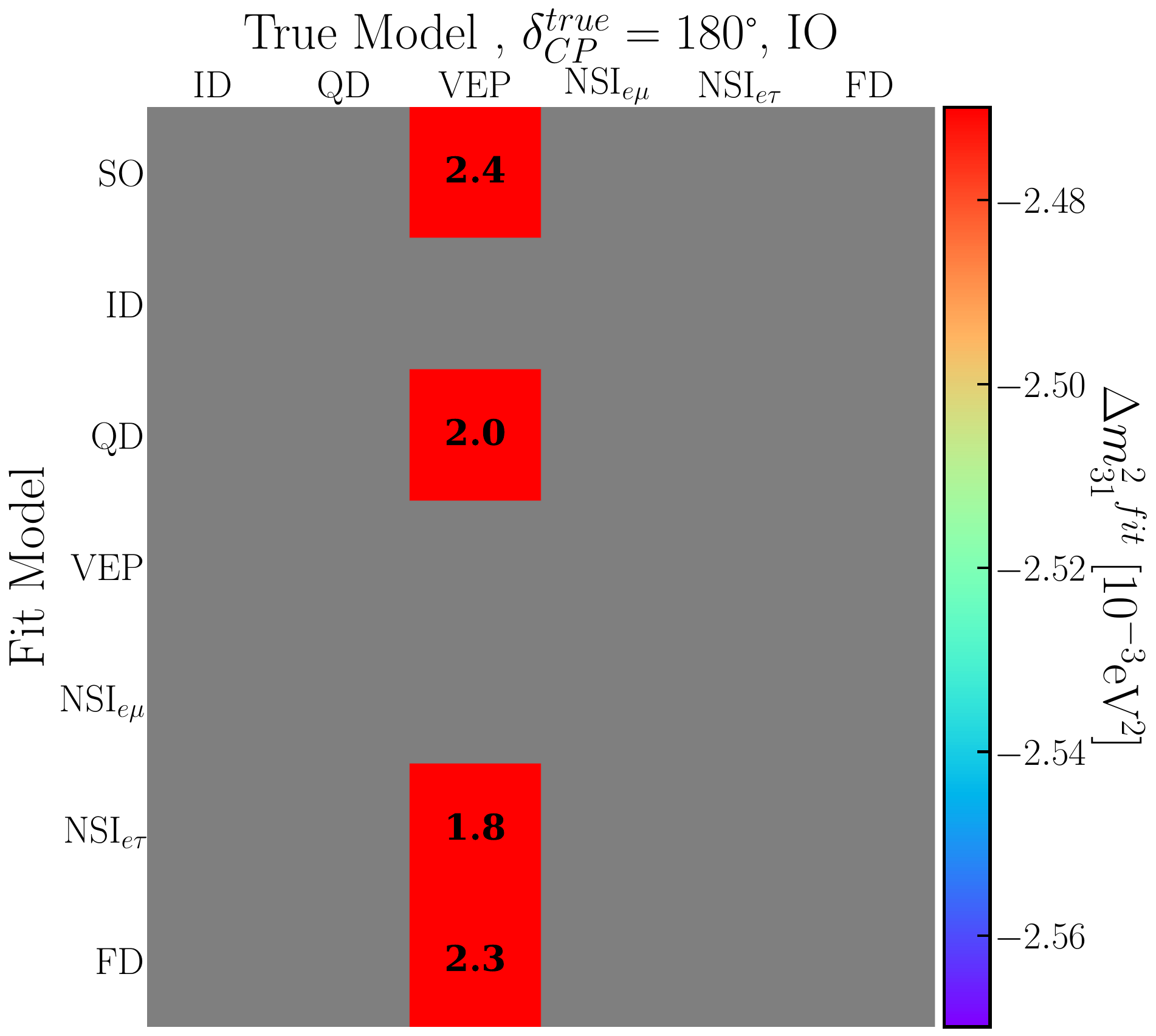} \\
  \end{tabular}
  \vspace{-0.3cm}
	\caption{Table of True Models vs Fit Models in which the values of $\Delta m^{2}_{31} {}^{fit}$ are when the deviation respect to $\Delta m^{2}_{31} {}^{true}$ is around $2.4 \sigma$. The boxes displaying numbers indicate the respective $N_\sigma$.}
 \label{A-Cuadro_dm31_3_sigma_dcp_-90_180}
\end{table}

\vspace{-0.2cm}

\subsection{Findings overview}
Below, we present a set of tables that will help us put into perspective the numerous results shown in the preceding section. These tables provide a panoramic and comparative overview, facilitating the drawing of general conclusions.

In each box, our tables pair a given true model (simulated data) with a fit model (theoretical hypothesis). All boxes in the tables are color-coded to highlight the quantity of interest related to the aforementioned pair analysis.

\subsubsection{ \texorpdfstring{$N_\sigma$} - summary}
The Table~\ref{A-Cuadro_3_sigma_dcp_-90_180} is built using all the results from our $N_\sigma$ analysis. The boxes of the tables corresponding to specific true and fit model combinations are color-coded to indicate the $\xi$ value at which a given fit model shows a 3$\sigma$ discrepancy from the true model. If this value of $\xi$ is below 
the upper limit of our range 
$(0.05)$, it is clear that there is still room for achieving $N_\sigma$ higher than 3$\sigma$, that would correspond to higher values of $\xi$. As we have seen from our previous plots, the pair values of $N_\sigma$ and $\xi$ are connected by an almost linear relationship. The gray boxes represent cases where the statistical separation between the given true model and the fit model is below the 3$\sigma$ level.

A particular note: the numbers in some of the boxes indicate the values of $N_\sigma$ for cases of initial discontinuities, where part of the curve is missing from the starting point to a specific point. For these special cases, the $N_\sigma$ separation for a given box could start from values higher than 3$\sigma$.

The model exhibiting an excellent discriminating power (from a theoretical hypothesis) is VEP, with NO and $\delta_{CP}^{true} = -90^\circ$, achieving a separation of 3$\sigma$ (or greater for the boxes with numbers) for SO, ID, QD and NSI$_{e\tau}$, with values of $\xi \gtrsim 0.020 \, (\Phi \Delta \gamma_{21} \gtrsim 1.15\times 10^{-24}), 0.022 \, (\Phi \Delta \gamma_{21} \gtrsim 1.30\times 10^{-24})$, and $0.029 \, (\Phi \Delta \gamma_{21} \gtrsim 1.74\times 10^{-24})$, respectively. For QD the separation starts from 6.4$\sigma$ 
for $\xi \gtrsim 0.035 \,(\Phi \Delta \gamma_{21} \gtrsim 2.05\times 10^{-24})$. 

If FD is selected as the true model, assuming NO and $\delta_{CP}^{true} = -90^\circ$, it shows a lower discrimination power than VEP. However, FD reaches the best discriminating power for NO and $\delta_{CP}^{true} = 180^\circ$, such as SO, QD and NSI$_{e\tau}$ is a 3$\sigma$ separation or more for $\xi \gtrsim$ 0.011 ($\alpha^{vis}_3 \gtrsim 4.49\times 10^{-6}$ eV$^2$), $\xi \gtrsim$ 0.012 ($\alpha^{vis}_3 \gtrsim 4.90\times 10^{-6}$ eV$^2$) and $\xi \gtrsim 0.016$ ($\alpha^{vis}_3 \gtrsim 6.25\times 10^{-6}$ eV$^2$), respectively. For VEP, the possible discrepancy start at 3.2$\sigma$, for $\xi \gtrsim 0.013$ ($\alpha^{vis}_3 \gtrsim 5.19\times 10^{-6}$ eV$^2$).

For IO, VEP is the only true model able to discriminate from several theoretical hypotheses. In fact, for both values of $\delta_{CP}^{true}$, VEP can be separated from SO, QD, NSI$_{e\tau}$ an FD. Therefore, from an overall perspective, i.e., for all the different combinations of mass hierarchy and $\delta_{CP}^{true}$, VEP is the easiest true model to distinguish from the fit ones. The VEP discriminating capacity in $N_\sigma$ relies on its well-known energy-increasing behavior in the probability, see Eq.~(\ref{DEvep}) (you can also check~\cite{Hoefken}), which poses challenges when fitting within other scenarios.  

\subsubsection{ \texorpdfstring{$\delta_{CP}$}- summary }

In Table~\ref{A-Cuadro_dcp_3_sigma_dcp_-90_180}, the color coding in the boxes (excluding the gray ones) indicates the value of $\delta_{CP}^{fit}$ that deviates from the true value at which a given number of standard deviations ($\sigma$) is reached. This deviation is not fixed across all the boxes in the table, as most values range from $2.4\sigma$ to $3.0\sigma$, with one exception corresponding to $3.6\sigma$, which will be explained in the next paragraph.

For instance, the $2.43\sigma$ separation, the lowest bound of the interval, occurs for NSI$_{e\mu}$ (true) and QD(fit). The case of $3.0\sigma$ separation occurs for the pair FD(true)-SO(fit), FD(true)-QD(fit), and VEP(true)-NSI$_{e\tau}$(fit), with $\delta^{true}_{CP} = -90^\circ$ and NO assumed. For 
$\delta^{true}_{CP} = -90^\circ$ and IO, the pair NSI$_{e\tau}$(true)-VEP(fit) and the pair NSI$_{e\tau}$(true)-NSI$_{e\mu}$(fit) are the ones with 3$\sigma$ separation. For $\delta^{true}_{CP} = 180^\circ$ and NO, the pairs with 3$\sigma$ separation are: VEP(true)-NSI$_{e\mu}$(fit),  VEP(true)-NSI$_{e\tau}$(fit) and FD(true)-NSI$_{e\tau}$(fit), while for IO these are: NSI$_{e\tau}$(true)-ID(fit), VEP(true)-NSI$_{e\tau}$(fit) and NSI$_{e\tau}$(true)-FD(fit). There is a unique case of 3.6$\sigma$ separation for FD(true)-VEP(fit), which happens because the fitting theory curve(VEP) starts after the 3.0$\sigma$ deviation. The gray boxes represent the pairs where the deviation of $\delta^{true}_{CP}$ from $\delta^{fit}_{CP}$ is below the lower bound of the interval mentioned above. The number inside the boxes denotes $N_\sigma$ for the given pair.

Here, we can observe very challenging cases (from a statistical perspective) where a given pair (true and fit model) could have a very small $N_\sigma$ separation but a sizable deviation between $\delta^{fit}_{CP}$ and $\delta^{true}_{CP}$. The most extreme case is for the NSI$_{e\mu}$(true) and NSI$_{e\tau}$(fit) pair, for $\delta^{true}_{CP} = -90^\circ$ and IO, where $N_\sigma = 0.1\sigma$, while, contrastingly, the deviation between $\delta_{CP}$'s is $3\sigma$ with $\delta^{fit}_{CP} = -125.99^\circ$. The second notable case occurs when 
$N_\sigma = 0.2\sigma$, with the
$\delta_{CP}$'s deviation equal to $3\sigma$, for NSI$_{e\tau}$(true)-ID(FD)(fit) with $\delta^{fit}_{CP} = 206.43(206.45)^\circ$, for 
 $\delta^{true}_{CP} = 180^\circ$ and IO. There are other cases with $N_\sigma = 0.2\sigma$, but their $\delta_{CP}$'s separation is below $3\sigma$. 

There are several other cases where a similar contrast is maintained, with a low value of $N_\sigma < 1\sigma$ and a $\sigma$ separation between $\delta_{CP}$'s above $2.4\sigma$. The common factor in these cases is that NSI is involved either as the true model, the fit model, or both. The latter can be reasonably expected since the NSI scenarios include an extra CP phase $\phi_{\alpha\beta}$, providing an additional degree of freedom. Thus, when the other BSO hypotheses are fitted, there is room for distortion in their fitted (and unique) standard CP phase, which is, roughly speaking, trying to accomodate 
$\delta^{fit}_{CP}\rightarrow\phi_{\alpha\beta} + \delta^{true}_{CP}  $. Of course, the same happens when NSI hypotheses are compared, i.e., $\phi^{fit}_{e\tau} + \delta^{fit}_{CP} \rightarrow \phi_{e\mu} + \delta^{true}_{CP}  $.

On the other hand, there are cases where the true and fit model pairs are highly distinguishable statistically (high $N_\sigma$). In these types of cases, it would be expected that a relevant deviation between $\delta^{fit}_{CP}$ and $\delta^{true}_{CP}$ (above $2.4\sigma$) would be disregarded as a controversial result. It is pertinent to mention that similar tables to those presented earlier are found in \cite{Denton:2022pxt}, with the distinction that, in our case, when testing the ability to distinguish between models, we allow the true values to vary within a range of values for
$\xi$.

\subsubsection{ \texorpdfstring{$\theta_{23}$} - summary }
Similar to before, in Table~\ref{A-Cuadro_th23_3_sigma_dcp_-90_180}, the color coding in the boxes (excluding the gray ones) indicates the value of $\theta_{23}^{fit}$ that deviates from the true value. The latter deviation ranges from $2.4\sigma$ to $3.0\sigma$, with some cases exceeding $3.0\sigma$. As explained previously, these latter cases occur when the fitting theory begins to provide a solution after surpassing the $3.0\sigma$ separation threshold.

The lowest separation is $2.43\sigma$, which occurs for the pair FD(true)-ID(fit) at $\delta^{true}_{CP} = 180^\circ$ and IO, with $\theta_{23}^{fit} = 50.06^\circ$. The cases with the highest separation are: QD(true)-NSI$_{e\tau}$(fit) with $3.16\sigma$ at $\delta^{true}_{CP} = 180^\circ$ and NO, with $\theta_{23}^{fit} = 50.13^\circ$; QD(true)-FD(fit) with $4.11\sigma$ at $\delta^{true}_{CP} = -90^\circ$ and IO, with $\theta_{23}^{fit} = 50.51^\circ$; and QD(true)-ID(fit) with $4.69\sigma$ at $\delta^{true}_{CP} = 180^\circ$ and IO, with $\theta_{23}^{fit} = 50.66^\circ$. The remaining pairs have separations ranging from $2.45\sigma$ to $3.0\sigma$.

The cases with the highest separation, mentioned above, correspond to challenging scenarios, characterized by $N_\sigma \rightarrow 0$ but significant deviations in the fitted parameter $\theta_{23}^{fit}$ from the true value. Apart from these, there are numerous other cases where $N_\sigma \rightarrow 0$ and the separation in $\theta_{23}$ reaches $3.0\sigma$. In the majority of cases where noticeable distortions in $\theta_{23}$ appear, QD plays the role of the true scenario. The signature pattern of QD is to decrease the probability amplitude by a nearly constant amount as energy increases~\cite{Carpio-2}. Therefore, when fitted against the other BSO hypotheses—which naturally do not incorporate this behavior—they are forced to make non-negligible changes to $\sin^2\theta^{fit}_{23}$ (which modulates the probability amplitude) to match with the value of the true probability.

\subsubsection{ \texorpdfstring{$\Delta m^2_{31}$} - summary }
The color coding in Table~\ref{A-Cuadro_dm31_3_sigma_dcp_-90_180} is consistent with that used in the preceding tables. Most of the fitted $\Delta m^2_{31}$ deviation from its true value corresponds to $3.0\sigma$. The only exception below this threshold, which also represents the lowest separation, is the case FD(true)-VEP(fit), with $2.73\sigma$ at $\delta^{true}_{CP} = -90^\circ$ and NO. The highest separation cases are: VEP(true)-QD(fit) with $6.47\sigma$ at $\delta^{true}_{CP} = -90^\circ$ and NO, and VEP(true)-QD(fit) with $11.24\sigma$ at $\delta^{true}_{CP} = 180^\circ$ and NO.

Regarding the more challenging cases, we have VEP(true)-NSI$_{e\tau}$(fit) with $N\sigma=0.4\sigma$ at $\delta^{true}_{CP} = -90^\circ$ and NO, with $\Delta {m^2_{31}}^{fit} = 2.408 \times 10^{-3}$eV$^2$ for a $3\sigma$ separation. The next challenging scenarios range between $N_\sigma = 0.8\sigma$ and $1.3\sigma$. It should be noted that most combinations of true and fit scenarios that reach $3\sigma$ also reach $5\sigma$. The VEP true scenario clearly dominates the cases where a significant separation in $\Delta m^2_{31}$ is observed. This behavior is expected, as the term $\Delta m^2_{31} L/E_\nu$ in the BSO scenarios requires substantial adjustments to match the form shown in Eq.~(\ref{DEvep}), which corresponds to the VEP interference term.

\subsubsection{Contrasting DUNE's Sensitivity to BSO Scenarios with Other Future Experiments}

When considering the BSO discriminating power of DUNE in comparison to other future neutrino experiments, we note that the parameter values for VEP and QD that can be tested at DUNE exceed JUNO's sensitivity \cite{JUNO:2015zny,JUNO:2021ydg}, primarily due to the $LE$ and $L$ dependencies of VEP and QD sensitivity, respectively. Regarding NSI, DUNE probes flavor-changing neutral-current interactions with electrons, whereas reactor experiments like JUNO are sensitive to charged-current interactions involving up and down quarks~\cite{Farzan:2017xzy}. Moreover, DUNE demonstrates better sensitivity to FD compared to JUNO, though JUNO is more competitive in detecting ID (see Table 3 in~\cite{Arguelles:2022tki}).

A similar scenario holds for experiments like T2HK, which, while approaching DUNE's sensitivity for ID, is not capable of detecting other BSO scenarios if their true parameters lie within the studied range. Interestingly, due to its reduced sensitivity to BSO effects, T2HK can play a complementary role by helping constrain the true values of standard neutrino oscillation parameters, as discussed in~\cite{We-Gabriela-Ternes}. Although a detailed analysis is necessary for a more precise assessment, it is reasonable to expect that T2HKK's discriminating power~\cite{Panda:2022vdw}, based on its baseline and average neutrino energy, should be similar to DUNE's.

On the other hand, experiments like IceCube-Gen2~\cite{IceCube-Gen2:2020qha}, with their improved sensitivity to high-energy neutrinos, including those from astrophysical and atmospheric sources, should be able to discriminate between different BSO phenomenologies that depend on factors like $L$, $LE$, or $L/E$ in a parameter region beyond DUNE's reach. This also applies to the KM3NeT experiment~\cite{KM3NeT:2018wnd}, which shares similar high-energy capabilities.

\renewcommand{\arraystretch}{1}




\section{Conclusions}

From the six BSO hypotheses used as DUNE true (simulated) data, we found that VEP is the most distinguishable from the other BSO hypotheses. Actually, VEP can be discriminated, at the 3$\sigma$ level and higher, from SO, and QD, regardless of the neutrino mass hierarchy and the value of $\delta^{true}_{CP}$. VEP can also be differentiated from ID (NSI$_{e\tau}$ and FD) for NO (IO) and for both values of $\delta^{true}_{CP}$. The maximum discrimination sensitivity for VEP is achieved when the fit model is SO, with $\delta^{true}_{CP}=180^\circ$ and IO. In this case, $\Phi \Delta \gamma_{21} \gtrsim 8.23 \times 10^{-25}$ corresponds to the range of values that allow for $N_\sigma$ discrimination at a significance level of 3$\sigma$ and above. Besides, SO, the second best VEP sensitivity, when the fit model is a BSO hypothesis, corresponds to FD with $\Phi \Delta \gamma_{21} \gtrsim 8.78 \times 10^{-25}$
 and for $\delta^{true}_{CP}=180^\circ$ and IO.

 Out of the many BSO models considered, only one besides VEP demonstrates any discrimination power, albeit weaker than VEP, and this is FD. Similar to the latter, VEP can be distinguished from SO and QD regardless of the values of $\delta^{true}_{CP}$ used, but only for the NO hierarchy. It can also be distinguished from VEP for the same cases mentioned before, while the distinction between FD and NSI$_{e\tau}$ could only occur for the IO hierarchy and for $\delta^{true}_{CP} = 180^\circ$. When FD is compared against SO, the maximum range of sensitivity is attained with $\alpha^{vis}_3 \gtrsim 4.49 \times 10^{-6}$eV$^2$.
Now, when the fit model is a BSO hypothesis, the best sensitivity is for QD, with the range of FD parameters able to make the distinction (at the 3$\sigma$ level and above) given by $\alpha^{vis}_3 \gtrsim 4.90 \times 10^{-6}$eV$^2$. On the other hand, if ID, QD, or any of the two NSI hypotheses is present in nature, DUNE will be unable to unravel any of them from the other BSO alternatives studied here, unless the distinction reaches the 3$\sigma$ level or above.

It is worth mentioning that if we reduce the number of degrees of freedom considered in our analysis (i.e., fixing $\theta_{23}$, for example), the discrimination power for the different BSO hypotheses will increase.

Our analysis also measures the distortion between the fitted values of $\delta_{CP}$, $\theta_{23}$, and $\Delta m^2_{31}$ and their respective true values within the context of model comparisons. All these comparisons indicate that it is possible to observe a significant separation between the fitted values and their true counterparts. Notably, this separation occurs even when the models are very similar in the global comparison (measured in terms of $N_\sigma$), i.e., when they are separated by less than $2σ$. The common factor in the cases corresponding to $\delta_{CP}$ is the involvement of the NSI hypotheses, which include an extra CP phase. This opens the possibility of generating observable deviations of the fitted standard CP phase from its true values. The QD scenario dominates the cases where a significant deviation in the fitted $\theta_{23}$ from its true value is observed. This occurs because the QD hypothesis decreases the probability amplitude, requiring the fitted $\theta_{23}$ (i.e., $\sin^2 \theta_{23}$) to be significantly altered to compensate. Most cases corresponding to noticeable separations between the fitted and true $\Delta m^2_{31}$ are characterized by the presence of the VEP hypothesis. This happens due to the difficulty in fitting (for the other scenarios) the effect of the growing energy-dependent component introduced by VEP in the true interference term .

Therefore, in this paper, we present a comprehensive and detailed roadmap for assessing DUNE's capabilities to distinguish among BSO hypotheses, if present in nature, as well as between BSO hypotheses and SO. Equally important to the global model separation is the mapping of potential distortions between the true and fitted values of $\delta_{CP}$, $\theta_{23}$, and $\Delta m^2_{31}$, studied individually for each parameter. The significance of our analysis is underscored by the importance of measurements such as $\delta_{CP}$ and $\theta_{23}$, which are expected to be pivotal achievements in neutrino physics in the coming years.

\section{Acknowledgements}

\noindent
This work is supported by Huiracocha Scholarship 2020 and {\it{Dirección
de Fomento de la Investigación}} at Pontificia Universidad
Católica del Perú, through Grants No. DFI-2021-0758 and CONCYTEC through Grant
No.060-2021-FONDECYT.


\bibliographystyle{utphys}
\bibliography{bibliography}

\end{document}